\documentclass[11pt,a4paper]{article}
\pdfoutput=1

\usepackage[utf8]{inputenc}
\usepackage[english]{babel}
\usepackage{chicago}

\usepackage[round]{natbib}
\usepackage[margin=1in]{geometry}

\usepackage{graphicx, subfig, amsmath, amssymb, ifsym}
\newtheorem{proposition}{Proposition}{\bfseries}{\itshape}

\usepackage{afterpage}

\usepackage{color}
\definecolor{blue}{rgb}{0.11,0.53,0.93}
\definecolor{red}{rgb}{0.8,0,0}

\usepackage{amsmath}
\newcommand{\x}{\boldsymbol{x}}
\newcommand{\z}{\boldsymbol{z}}

\newcommand{\X}{\boldsymbol{X}}

\newcommand{\A}{\boldsymbol{A}}

\newcommand{\D}{\boldsymbol{D}}

\newcommand{\M}{\boldsymbol{M}}
\newcommand{\MI}{\boldsymbol{M}_{\textsf{I}}}
\newcommand{\MII}{\boldsymbol{M}_{\textsf{II}}}
\newcommand{\SIR}{\textsf{SIR}}
\newcommand{\SAVE}{\textsf{SAVE}}
\newcommand{\LDA}{\textsf{LDA}}

\newcommand{\Sigmab}{\boldsymbol{\Sigma}}

\newcommand{\Omegab}{\boldsymbol{\Omega}}
\newcommand{\w}{\omega}
\newcommand{\I}{\boldsymbol{I}}
\newcommand{\betab}{\boldsymbol{\beta}}
\newcommand{\mub}{\boldsymbol{\mu}}

\newcommand{\0}{\boldsymbol{0}}
\newcommand{\T}{{}^{\top}}
\newcommand{\Space}{\mathcal{S}}

\DeclareMathOperator{\Real}{\mathbb{R}}

\renewcommand{\hat}[1]{\widehat{#1}}
\DeclareMathOperator*{\argmax}{arg\,max}
\DeclareMathOperator*{\diag}{diag}

\DeclareMathOperator{\Exp}{\mathrm{E}}
\DeclareMathOperator{\Var}{\mathrm{Var}}

\usepackage{mathrsfs}

\begin{document}


\title{Graphical Tools for Model-based Mixture Discriminant Analysis}
\author{Luca Scrucca \\ Universit\`a degli Studi di Perugia}
\date{\today}

\maketitle

\begin{abstract}
The paper introduces a methodology for visualizing on a dimension reduced subspace the classification structure and the geometric characteristics induced by an estimated Gaussian mixture model for discriminant analysis. In particular, we consider the case of mixture of mixture models with varying parametrization which allow for parsimonious models.
The approach is an extension of an existing work on reducing dimensionality for model-based clustering based on Gaussian mixtures. 
Information on the dimension reduction subspace is provided by the variation on class locations and, depending on the estimated mixture model, on the variation on class dispersions. 
Projections along the estimated directions provide summary plots which help to visualize the structure of the classes and their characteristics. 
A suitable modification of the method allows us to recover the most discriminant directions, i.e., those that show maximal separation among classes.
The approach is illustrated using simulated and real data.

\end{abstract}


\section{Introduction}

Discriminant analysis or supervised learning indicates a broad set of statistical methods aimed at classifying a categorical outcome variable $Y$, an indicator with $K$ classes, on the basis of a $(p \times 1)$ vector of features $\x$. 
Among the several methods available for continuous 
features, one of the most popular approach is classical linear discriminant analysis (LDA). This method has been extended to quadratic discriminant analysis (QDA), and, more generally, to models based on finite mixture modeling of Gaussian densities. 

Independently from the statistical method adopted, visualization and graphics can play an important role in the understanding of the classification results. 
Typically, canonical variates are computed when the dimension of the features space is large. This allows us to visualize the classes on a reduced subspace, often bi-dimensional. 
However, canonical variates are tailored to LDA. Some methods have been proposed for QDA, while graphical methods for finite mixture modeling is still a research area to be explored.
From a different point of view, \cite{Hennig:2004} has proposed an asymmetric discriminant projection method by looking at the projections where a class appears as homogeneous as possible and separated from the remaining groups.

In this paper a dimension reduction method for visualizing and summarizing the fit of a model-based mixture discriminant analysis is discussed. 
The approach is an extension of the method proposed by \cite{Scrucca:2010} for model-based clustering. 
The estimated subspace is found by looking at the variation in class means and class covariances depending upon the assumed parameterization of the fitted Gaussian mixture model. 
The resulting projection subspace is able to capture most of the classification structure available in the data. 
The proposal reduces to LDA canonical variates for a specific parameterization of the mixture model, while it is equivalent to a recently proposed method for QDA. 
In all the other cases, the proposed visualization method is able to show the main geometric characteristics of the fitted mixture model. 
Furthermore, the proposal can be adapted to allow for the visualization of the separation among the classes.

The paper is organized as follows. 
In Section 2 a brief review of classification and graphical methods based on the Gaussian distribution is provided. Section 3 describes the Gaussian mixture models for discriminant analysis we consider in this paper. 
In Section 4 the methodology is introduced and the main properties are described. 
Section 5 presents examples based on both simulated and real data. 
In Section 6 the proposed method is extended to allow to recover the most discriminant directions, i.e., those that show maximal separation among classes.
Concluding remarks appear in the final Section.


\section{Classification based on the Gaussian distribution and existing graphical methods}

All the models discussed in this paper are probabilistic, i.e., based on the assumption that the observations in the $k$th class ($k=1,\ldots,K$) are generated by a probability distribution $f_k(\x)$, where $\x = (x_1, x_2, \ldots, x_p)\T$ is a column vector of $p$ observed features.
Most discriminant analysis methods for continuous variables are based on the assumption that observations in each class are multivariate normal, so that $f_k(\x) = \phi(\x|\mub_k, \Sigmab_k)$, where $\phi$ is the $p$-variate Gaussian density with mean $\mub_k$ and covariance matrix $\Sigmab_k$. 

Linear discriminant analysis (LDA) assumes normal populations with equal class covariance matrices, i.e., $\Sigmab_k = \Sigmab_W$ for all $k=1,\ldots,K$, where $\Sigmab_W = \sum_{i=k}^K \pi_k \Sigmab_k$ is the within-class covariance matrix with class prior probabilities $\pi_k$. 
The resulting discriminant function is linear in the feature vector $\x$ and the acceptance regions for the classes are separated in $\Real^p$ by means of hyperplanes.
However, Fisher's \citeyearpar{Fisher:1936} original proposal did not rely on the Gaussian distribution. Based on geometric arguments, he looked for a vector of $d$ linear combinations $\betab\T\x$, with $\betab \in \Real^{p \times d}$, such that the between-class covariance, 
$\Sigmab_B = \sum_{k=1}^K \pi_k (\mub_k - \mub)(\mub_k - \mub)\T$ with $\mub = \sum_{k=1}^K \pi_k \mub_k$,
is maximized relative to the within-class covariance, $\Sigmab_W$. 
This amounts to maximize the so called \textit{Rayleigh quotient}, i.e.,
$$
\argmax_{\betab}\;
\frac{\betab\T \Sigmab_B \betab}{\betab\T \Sigmab_W \betab},
$$
or, equivalently, find $\betab \in \Real^{p \times d}$ which maximizes $\betab\T \Sigmab_B \betab$ subject to $\betab\T \Sigmab_W \betab = \I_d$, where $\I_d$ is the identity matrix of dimension $(d \times d)$. 
The problem is solved by the generalized eigenvalue decomposition of $\Sigmab_B$ with respect to $\Sigmab_W$.

The directions given by the $d$ columns of $\betab$ form the basis of the $d$-dimensional reduction subspace, $\Space(\betab)$, which shows the maximal separation among classes, and decision boundaries are linear in the projected features subspace. The dimension of this subspace is $d = \min(p,K-1)$, so just one direction can be estimated in two-class problems.
Fisher's or LDA canonical variates, $\betab\T\x$, express the projection onto this subspace, and provide a graphical counterpart to LDA \cite[Chap. 11]{Mardia:Kent:Bibby:1979}.

There exists some connection between LDA canonical variates and other dimension reduction methods. In particular, it has been shown that for a categorical response variable sliced inverse regression \cite[SIR;][]{Li:1991} is equivalent to LDA, except for a different scaling. In fact, SIR covariates are scaled to have unit covariance while the LDA canonical variates are scaled to have unit within-class covariance \citep{Chen:Li:2001}. See also \cite{Kent:1991} for more discussion on the connection between SIR and LDA. 

Quadratic discriminant analysis (QDA) is obtained by removing the assumption of a common class covariance matrix. The resulting discriminant function is quadratic, and the decision boundaries are quadratic surfaces over the features subspace. However, in this case there appears to be no standard canonical variates analysis for QDA as there is for LDA. 
Some authors have considered dimension reduction methods for quadratic discrimination in normal populations with different covariance matrices. \cite{Pardoe:Yin:Cook:2007} proposed the use of sliced average variance estimation \citep[SAVE;][]{Cook:Weisberg:1991} as a graphical representation in quadratic discriminant analysis. \cite{Velilla:2008, Velilla:2010} discussed the concept of quadratic subspace as a tool for dimension reduction in QDA. 

Other extensions to LDA are regularized discriminant analysis \cite[RDA,][]{Friedman:1989}, flexible discriminant analysis \cite[FDA,][]{Hastie:Tibshirani:Buja:1994}, and penalized discriminant analysis \cite[PDA,][]{Hastie:Buja:Tibshirani:1995}.
RDA represents a compromise between LDA and QDA; it uses a tuning parameter $\alpha$ for class covariance matrix estimation, i.e., the covariance matrix for class $k$ is estimated by the convex combination 
$$
\Sigmab_k(\alpha) = \alpha\Sigmab_k + (1-\alpha)\Sigmab_W.
$$
FDA fits by optimal scoring a linear regression model using a basis expansion $h(\x)$ of the feature vector $\x$.
PDA also uses optimal scoring on a basis expansion $h(\x)$ as in FDA, but with a quadratic penalty on the coefficients, i.e., solving the following optimization problem 
$$
\argmax_{\betab}\;
\betab\T \Sigmab_B \betab
\quad \text{subject to } 
\betab\T (\Sigmab_W + \lambda\Omegab) \betab = \I_d,
$$
where $\Omegab$ is a $(p \times p)$ symmetric, nonnegative definite, penalty matrix.
All these methods have no direct graphical representation associated, so usually canonical variates are computed as in LDA using the estimated class means.

Finally, we mention the LAD proposal by \citet{Cook:Forzani:2009}, a likelihood-based dimension reduction method under the assumption of conditional normality of predictors given the response. This model closely resembles the family of models we adopted, but the estimation procedure is quite different. In fact, no closed-form solution to the maximum likelihood estimation of the central subspace (the parameter of interest) is available. Thus, numerical optimization is used for maximization of the log-likelihood on Grassman manifolds.

\section{Finite mixture modelling in discriminant analysis}
\label{sec:mixmodclassif}

Mixture discriminant analysis generalizes the previous approaches by allowing the density for each class conditional density to be expressed by a finite mixture of normals. Thus, a Gaussian mixture model for the $k$-th class ($k=1,\ldots,K$) has density
\begin{equation}
f_k(\x) = \sum_{g=1}^{G_k} \pi_{gk} \phi(\x; \mub_{gk}, \Sigmab_{gk}),
\label{eq:gaussian:mixture}
\end{equation}
where $\pi_{gk}$ are the mixing probabilities ($\pi_{gk} > 0$, $\sum_{g=1}^{G_k}\pi_{gk}=1$), $\mub_{gk}$ is the mean of component $g$ in class $k$, and $\Sigmab_{gk}$ is the covariance matrix of component $g$ in class $k$. 
Thus, Gaussian components are ellipsoidal, centered at $\mub_{gk}$, and with other geometric features, such as volume, shape and orientation, determined by $\Sigmab_{gk}$. 

\citet{Hastie:Tibshirani:1996} introduced mixture discriminant analysis (MDA) assuming a common full covariance matrix, i.e. $\Sigmab_{gk} = \Sigmab$ for all $g,k$, with known fixed number of mixture components for each class.

In a procedure named eigenvalue decomposition discriminant analysis (EDDA), \citet{Bensmail:Celeux:1996} proposed the use of Gaussian finite mixture modeling for discriminant analysis in which each class is modeled by a single Gaussian term, i.e., $G_k = 1$ for all $k$, with the same (possibly parsimonious) class covariance structure factorized as
$$
\Sigmab_{k} = \lambda_k\D_k\A_k\D\T_k,
$$
where
$\lambda_{k}$ is a scalar value controlling the volume of the ellipsoid, 
$\A_{k}$ is a diagonal matrix specifying the shape of the density contours, and
$\D_{k}$ is an orthogonal matrix which determines the orientation of the ellipsoid
\citep{Banfield:Raftery:1993, Celeux:Govaert:1995}.
Table~\ref{tab:mclust} shows the MCLUST family of mixture models supported by the \texttt{mclust} package \citep{Fraley:Raftery:Murphy:Scrucca:2012} for the \textsf{R} software \citep{RStatSoft}.

\begin{table*}[htb]
\caption{Parametrizations of covariance matrices available in the \texttt{mclust} software \citep{Fraley:Raftery:Murphy:Scrucca:2012} and related geometric characteristics.}
\label{tab:mclust}
\centering
\begin{tabular}{lccccc}
\hline\noalign{\smallskip}
Model & $\Sigmab_k$ & Distribution & Volume & Shape & Orientation \\
\noalign{\smallskip}\hline\noalign{\smallskip}
E   & $\sigma$          & Univariate & equal & & \\
V   & $\sigma_k$        & Univariate & variable & &  \\
EII & $\lambda\I$       & Spherical & equal & equal & \\
VII & $\lambda_k\I$     & Spherical & variable & equal & \\
EEI & $\lambda\A$       & Diagonal & equal & equal & coordinate axes \\
VEI & $\lambda_k\A$     & Diagonal & variable & equal & coordinate axes \\
EVI & $\lambda\A_k$     & Diagonal & equal & variable & coordinate axes \\
VVI & $\lambda_k\A_k$   & Diagonal & variable & variable & coordinate axes \\
EEE & $\lambda\D\A\D\T$ & Ellipsoidal & equal & equal & equal \\
EEV & $\lambda\D_k\A\D\T_k$ & Ellipsoidal & equal & equal & variable \\
VEV & $\lambda_k\D_k\A\D\T_k$ & Ellipsoidal & variable & equal & variable \\
VVV & $\lambda_k\D_k\A_k\D\T_k$ & Ellipsoidal & variable & variable & variable \\
\noalign{\smallskip}\hline
\end{tabular}
\end{table*}

A generalization of the previous two approaches is the method called MclustDA \citep{Fraley:Raftery:2002}, where a density estimate for the data is obtained by a Gaussian finite mixture model with a different number of components and a different (possibly parsimonious) covariance matrix for each class. The corresponding family is thus very flexible allowing the distribution of each class to be approximated by a mixture of Gaussian components.

Maximum likelihood estimates for finite mixture models can be computed via the EM algorithm \citep{Dempster:Laird:Rubin:1977}. Model selection, which requires choosing the number of mixture components and the covariance parameterization for each class, is usually based on penalized criteria, such as the Bayesian information criterion \cite[BIC,][]{Schwartz:1978} or the integrated complete likelihood \cite[ICL,][]{Biernacki:Celeux:Govaert:2000}.

\section{Dimension reduction for model-based discriminant analysis}
\label{sec:drmbda}

\subsection{Methodology}

Suppose we would like to find a suitable reduced number of projections which, depending on the estimated Gaussian mixture model, are able to visualize variation both in groups location and dispersion. Following the proposal of \cite{Scrucca:2010} for model-based clustering, consider the following matrices:
$$
\MI = \sum_{k=1}^K \sum_{g=1}^{G_k} \w_{gk} (\mub_{gk} - \mub)(\mub_{gk} - \mub)\T,
$$
where 
$\w_{gk} = \pi_k\pi_{gk}$ ($\w_{gk} > 0$, $\sum_{k,g} \w_{gk} = 1$),
$\mub = \sum_{k=1}^K \pi_k \mub_k = \sum_{k,g} \w_{gk} \mub_{gk}$ is the marginal mean vector, 
$\mub_k = \sum_{g=1}^{G_k} \pi_{gk}\mub_{gk}$ is the mean vector for class $k$, and
$$
\MII = \sum_{k=1}^K \sum_{g=1}^{G_k} \w_{gk} (\Sigmab_{gk} - \bar{\Sigmab}) \Sigmab_X^{-1} (\Sigmab_{gk} - \bar{\Sigmab})\T,
$$
where 
$\bar{\Sigmab} = \sum_{k,g} \w_{gk} \Sigmab_{gk}$ 
is the pooled within-class covariance matrix, and 
$\Sigmab_X = \frac{1}{n} \sum_{i=1}^{n} (\x_i - \mub)(\x_i - \mub)\T$ 
is the marginal covariance matrix.

The matrix $\MI$ contains information on class-component means variation, while $\MII$ contains information on class-component covariances variation. The two types of information can be summarized using the following kernel matrix
\begin{equation}
\M = \MI \Sigmab_X^{-1} \MI + \MII.
\label{GMMDRC:kernel}
\end{equation}
The matrix $\betab \in \Real^{p \times d}$ $(d=\min(p,\sum_{k=1}^{K}G_k -1)$ spanning the desired subspace is the solution of the following optimization
\begin{equation}
\argmax_{\betab}\;
\betab\T \M \betab 
\qquad \text{subject to} \quad
\betab\T \Sigmab_X \betab = \I_d,
\label{GMMDRC:optimization}
\end{equation}
where $\I_d$ is the $(d \times d)$ identity matrix.
The solution of \eqref{GMMDRC:optimization} is obtained through the generalized eigen-decomposition of $\M$ with respect to $\Sigmab_X$. Hence, the basis $\betab$ of the dimension reduction subspace $\Space(\betab)$ is obtained as $\Sigmab_X^{-1/2}$ times the eigenvectors of $\Sigmab_X^{-1/2}\M\Sigmab_X^{-1/2}$, with directions ordered according to the corresponding eigenvalues.
The projections onto such subspace is then given by $\z = \betab\T\x$. 
In analogy with the name of the method proposed in \cite{Scrucca:2010}, these are called GMMDRC (Gaussian Mixture Model Dimension Reduction for Classification) variables and provide a graphical method to display the classification structure resulting from a Gaussian mixture model. 

Note that in the presentation made so far we have assumed that the parameters of the population are known. Usually, however, they are unknown and must be estimated from training data as discussed in Section~\ref{GMMDRC:estimation}.

\subsection{Properties}

For MDA models the subspace spanned by $\M$ is equivalent to that spanned by $\MI$. This because under the MDA assumption of common class covariance, i.e., $\Sigmab_{gk} = \Sigmab$ for all $g,k$, we get $\MII = 0$, so no contribution comes from the variation on class covariances.
The same also happens for those EDDA models which assume constant class covariance matrices (i.e., models EII, EEI, and EEE -- see Table~\ref{tab:mclust}), because here $\Sigmab_{k} = \Sigmab$ for all $k$.
In all the other cases, i.e., for those mixture models which allow different class covariance matrices, $\MII$ adds further information for the identification of the reduction subspace.

In two specific cases the subspace identified by GMMDRC reduces to simple known situations as described in the following propositions, whose proofs are contained in the Appendix.

\begin{proposition}
\label{prop1}
Consider the EDDA mixture model with common full class covariance matrix (EEE). The subspace $\Space(\betab)$, obtained by solving the GMMDRC constrained optimization in \eqref{GMMDRC:optimization} with $\M = \MI \Sigmab_X^{-1} \MI$ , is identical to the subspace $\Space(\betab^\SIR)$ spanned by SIR, and also to the subspace $\Space(\betab^\LDA)$ spanned by LDA canonical directions.
\end{proposition}

Based on Proposition~\ref{prop1} we claim that using canonical LDA variables is only relevant when the adopted mixture model for classification assumes a single component with common covariance matrix for each class \cite[see also][]{Chen:Li:2001}.

\begin{proposition}
\label{prop2}
Consider the EDDA mixture model with different full class covariance matrices (VVV). The subspace $\Space(\betab)$, obtained by solving the GMMDRC constrained optimization in \eqref{GMMDRC:optimization} with $\M$ as in \eqref{GMMDRC:kernel}, is identical to the subspace $\Space(\betab^\SAVE)$ spanned by SAVE.
\end{proposition}

Noting that an EDDA mixture model with a different full covariance matrix for each class is essentially equivalent to QDA, Proposition~\ref{prop2} supports the use of SAVE as a graphical counterpart to QDA.

\begin{proposition}
\label{prop3}
Let $l_1 \ge l_2 \ge \dots \ge l_d > 0$ be the eigenvalues from the generalized eigen-decomposition of the kernel $\M$, i.e., $\M\betab_j = l_j\Sigmab_X\betab_j$, for $j=1,\ldots,d$. 
Each eigenvalue $l_j$, corresponding to the direction $\betab_j$ of the projection subspace $\Space(\betab)$, can be written as
\begin{equation}
l_j = \Var(\Exp(\betab\T_j\x|Y))^2 + \Exp(\Var(\betab\T_j\x|Y)^2)
\qquad \text{for }j=1,\ldots,d.
\label{eq:eigenvalues}
\end{equation}
Thus, the eigenvalues can be decomposed in the sum of the contributions given by 
\begin{itemize}
\item the squared variance of the between class-component means,
\item the average of the squared within class-component variances,
\end{itemize}
along the corresponding directions.
\end{proposition}

This result provides an interpretation for the contribution of each direction to the visualization of the classification structure. Along each GMMDRC direction classes can be separated by location, by dispersion, or both.
In addition, those directions associated with zero or approximately zero eigenvalues can be neglected since their contribution to class location or dispersion is negligible.
A formal assessment of dimensionality could be pursued, for instance, by implementing a permutation test as described in \cite{Li:1991}, however, is beyond the scope of this paper and deserves further study and investigation.

\subsection{Estimation}
\label{GMMDRC:estimation}

For a $(n \times p)$ sample data matrix $\X$ and the corresponding $(n \times 1)$ vector $Y$ containing the observed classes, the sample version $\hat{\M}$ of \eqref{GMMDRC:kernel} is obtained using the corresponding estimates from the fit of a Gaussian finite mixture model among those discussed in Section~\ref{sec:mixmodclassif}. 
Then, sample GMMDRC directions are calculated from the generalized eigen-decomposition of $\hat{\M}$ with respect to $\hat{\Sigmab}_X$, the sample marginal covariance matrix.

\section{Examples}

\subsection{Waveform data}

This is an artificial three-class problem with $p = 21$ variables, often used in machine learning and considered to be a difficult pattern recognition problem \citep{Breiman:etal:1984, Hastie:Tibshirani:1996}.
Consider the following three shifted triangular waveforms defined as
$$
w_1(j) = \max(6-|j- 11|,0), \quad
w_2(j) = w_1(j-4), \quad
w_3(j) = w_1(j+4),
$$
for $j=1,\ldots,21$.
Then, the variables $X_j$ are generated within each class $Y$ as a random convex combination of two basic waveforms with noise added:
\begin{equation*}
X_j = \begin{cases}
u_1 w_1(j) + (1-u_1) w_2(j) + \epsilon_j & \text{for } Y = 1 \\ 
u_2 w_2(j) + (1-u_2) w_3(j) + \epsilon_j & \text{for } Y = 2 \\
u_3 w_3(j) + (1-u_3) w_1(j) + \epsilon_j & \text{for } Y = 3
\end{cases}\; ,
\end{equation*}
where $j = 1,2,\ldots,21$, $w_h = (w_h(1), \ldots, w_h(21))\T$ for $h=1,2,3$, $(u_1, u_2, u_3)$ be independent random variables uniformly distributed on $[0,1]$, and $\epsilon_j$ following a standard normal distribution. 

\begin{figure}[htpb]
\subfloat[]
  {\includegraphics[width = 0.5\linewidth]{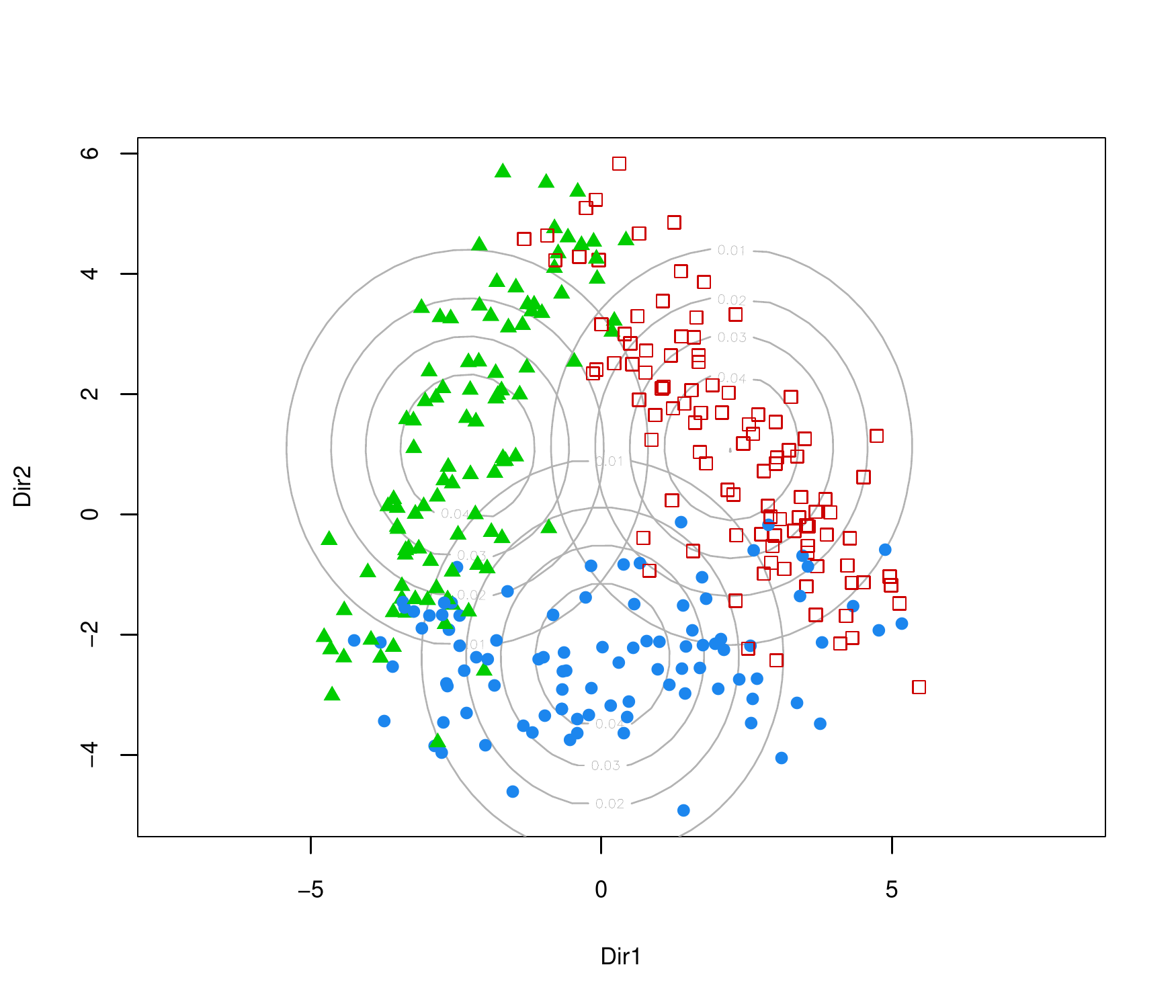}}
\subfloat[]
  {\includegraphics[width = 0.5\linewidth]{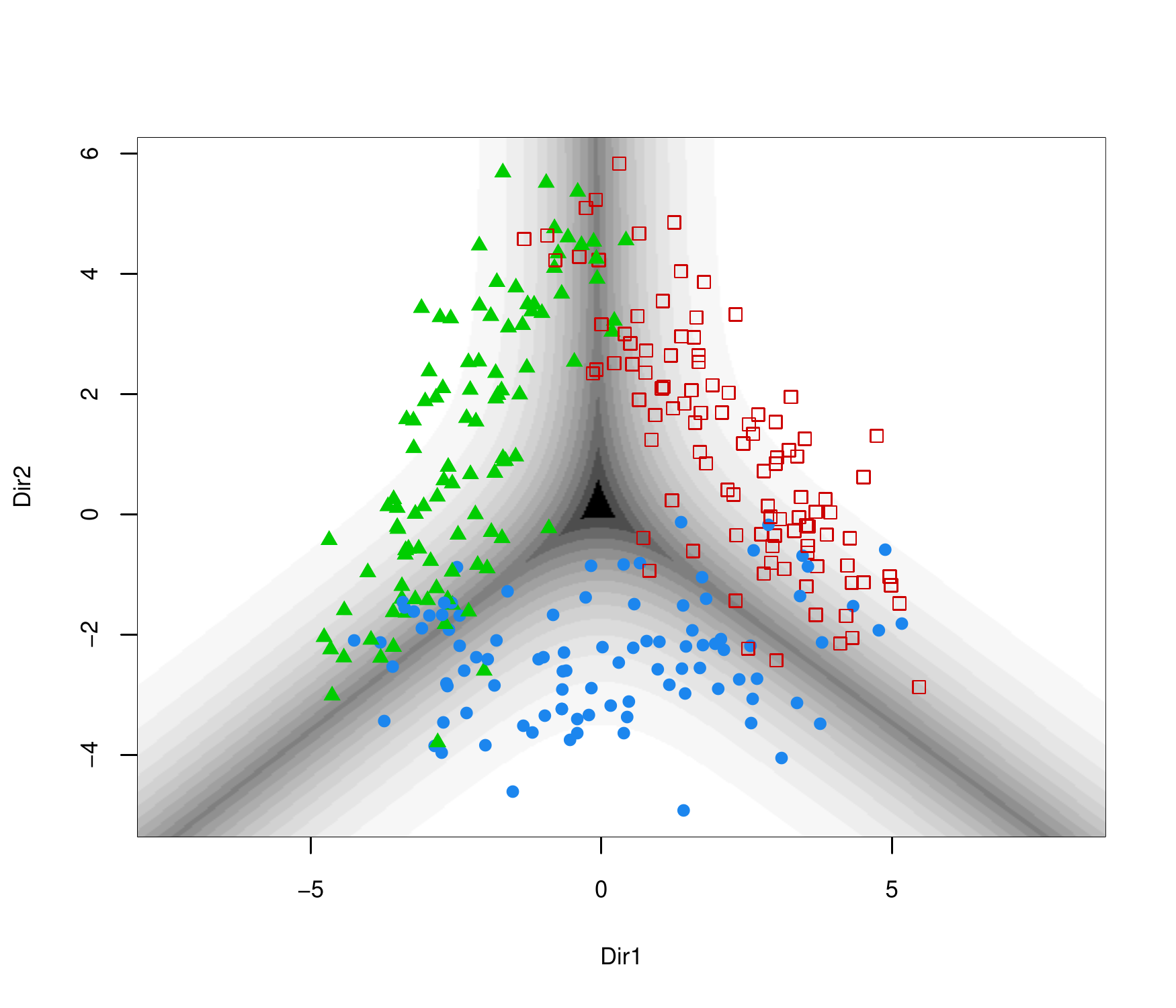}}
\caption{Plot of waveform data projected onto the first two directions for the EDDA model with common class covariance. Panel (a) shows the class density contours, while panel (b) the decision boundary with corresponding uncertainty.}
\label{fig1:waveform}
\end{figure}

Figure~\ref{fig1:waveform} shows a scatterplot of data points projected onto the directions estimated for the EDDA mixture model with EEE covariance structure, i.e., assuming a common class covariance. As already mentioned, this is equivalent to a plot of LDA canonical variates. 
Panel (a) contains the density contours for the three classes, which have the same shape, orientation, and volume. 
The graph on panel (b) displays the corresponding decision boundaries with associated classification uncertainty, with the uncertainty shown using a greyscale where darker regions indicating higher uncertainty. As expected the decision boundaries are linear.

\begin{figure}[htpb]
\subfloat[]
  {\includegraphics[width = 0.5\linewidth]{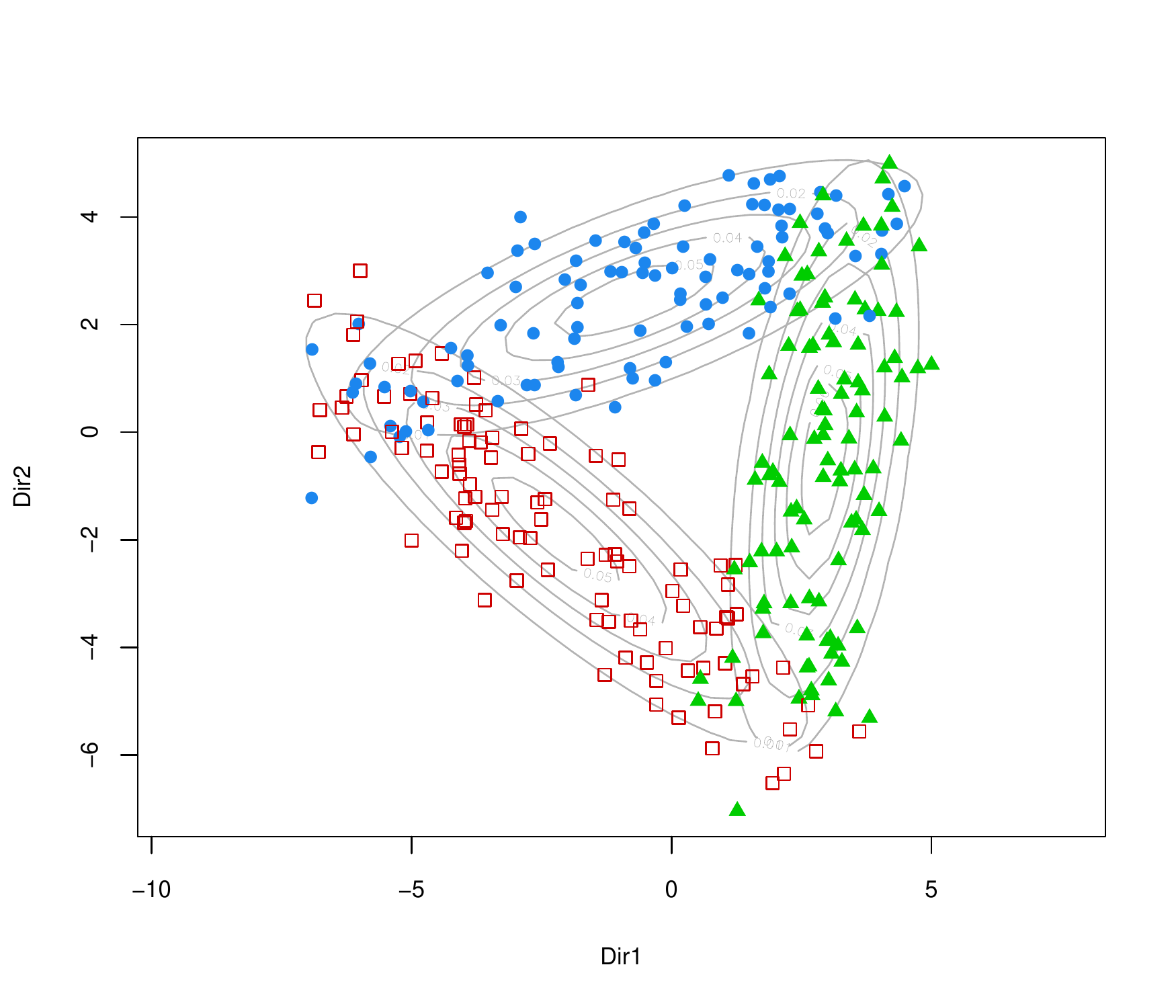}}
\subfloat[]
  {\includegraphics[width = 0.5\linewidth]{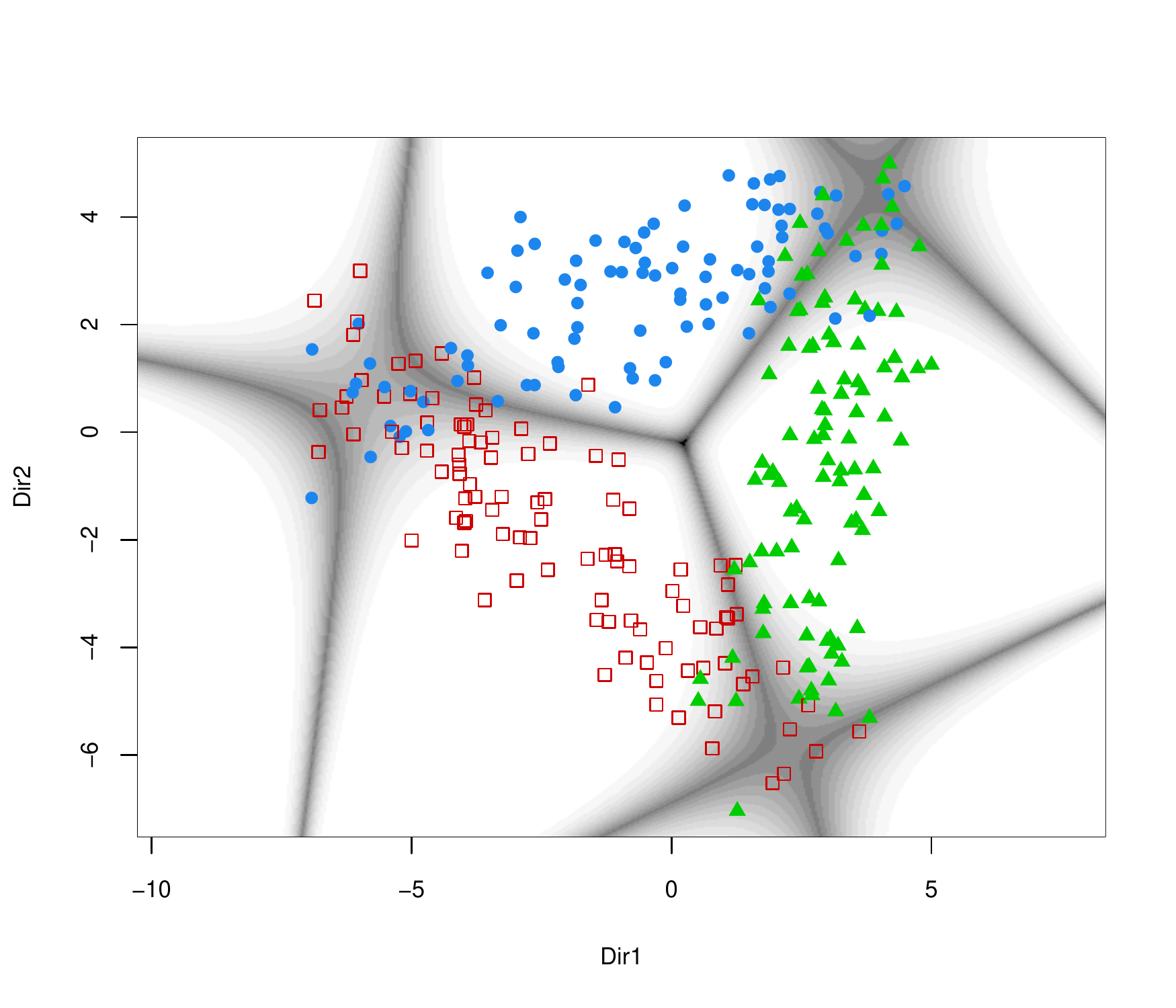}}
\caption{Plot of waveform data projected onto the first two directions for the EDDA model with different class covariances. Panel (a) shows the class density contours, while panel (b) the decision boundary with corresponding uncertainty.}
\label{fig2:waveform}
\end{figure}

Moving to a more complex model, we fitted an EDDA mixture model with VVV covariance structure, i.e., different class covariances. The corresponding projection is shown in Figure~\ref{fig2:waveform}. In this case, the contours have different orientation (see panel (a)) because no restrictions were placed on the class covariance matrices, hence the estimated model provides a better approximation to the data distribution. 
The triangular form of the data appears more clearly than in the previous case. 
The plot on panel (b) contains the classification boundaries given by quadratic polygons, which shows a lower overall uncertainty than in the previous case. 

\begin{figure}[htbp]
\subfloat[]
  {\includegraphics[width = 0.5\linewidth]{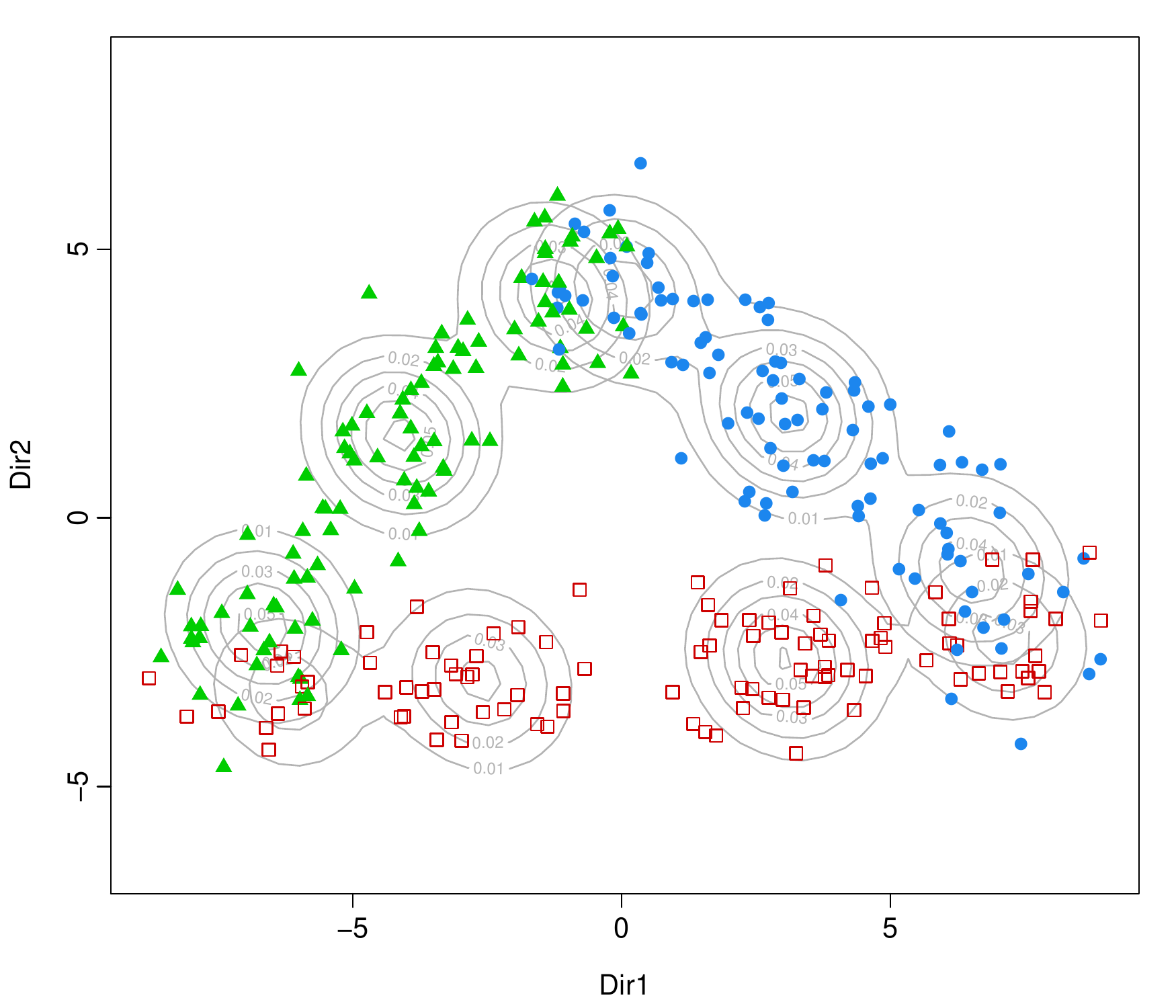}}
\subfloat[]
  {\includegraphics[width = 0.5\linewidth]{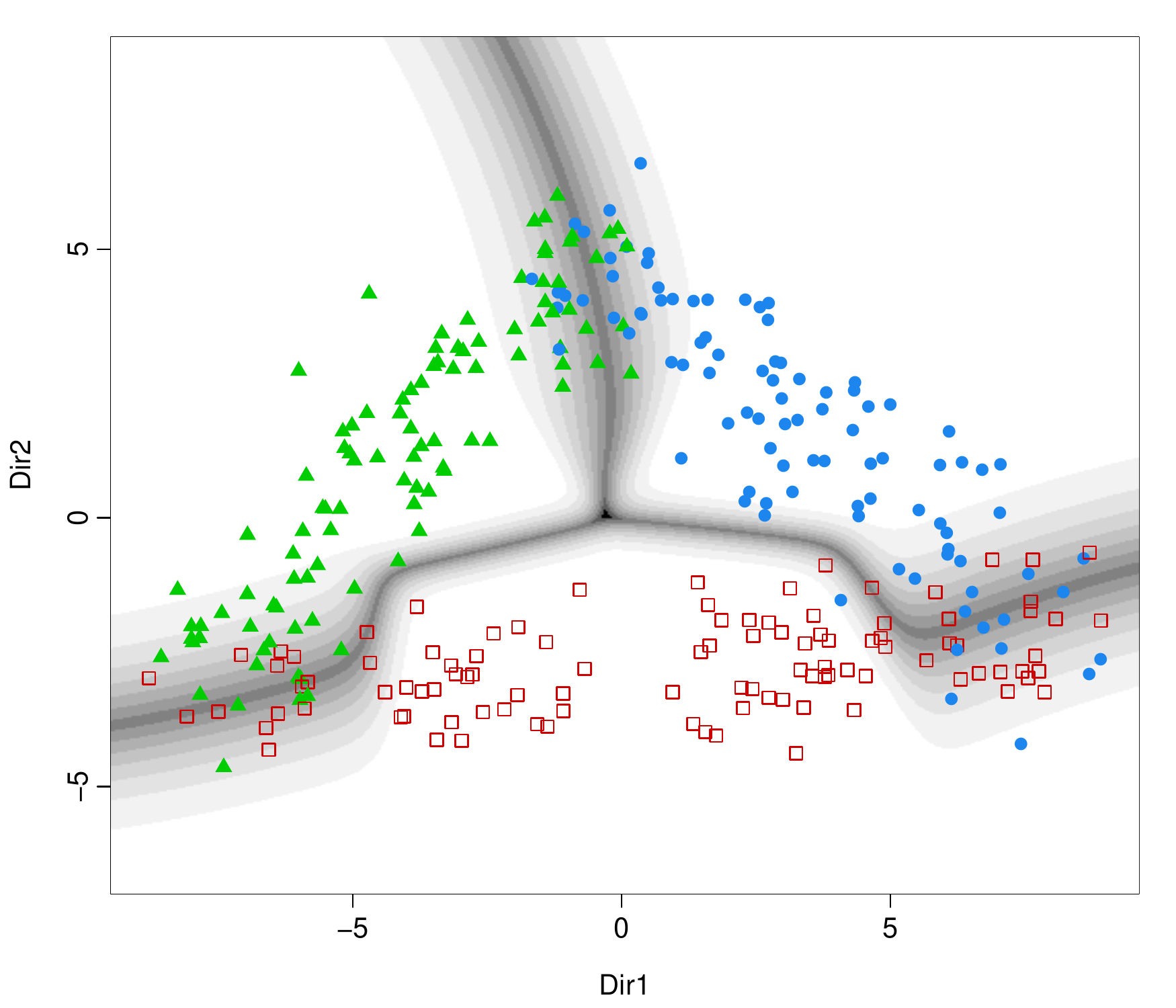}}
\caption{Plot of waveform data projected onto the first two directions for the selected MclustDA model. Panel (a) shows the class density contours, while panel (b) the decision boundary with corresponding uncertainty.}
\label{fig3:waveform}
\end{figure}

Finally, Figure~\ref{fig3:waveform} shows the data projected onto the first two directions estimated from the selected MclustDA mixture model. 
This model fitted a mixture of three class-specific Gaussian mixtures where the class-specific mixtures had $G_1=3$, $G_2=4$, and $G_3=3$ spherical Gaussian distributions as components.
These characteristics are clearly visible on panel (a) of Figure~\ref{fig3:waveform}. The resulting decision boundaries are shown on Figure~\ref{fig3:waveform}b; these appear to be highly nonlinear with a relative larger uncertainty where the classes overlap.
Finally, note that, on the basis of the corresponding eigenvalues, the first two directions account for 96\% of the overall information available in the 21 estimated directions.

\subsection{Swiss banknotes}

\citet[Table 1.1 and 1.2]{Flury:Riedwyl:1988} presented a dataset containing six physical measurements made on a sample of 1000 Swiss Franc bills. 100 observations were classified as genuines, and 100 as counterfeits.

\begin{figure}[htb]
\subfloat[]
  {\includegraphics[width = 0.5\linewidth]{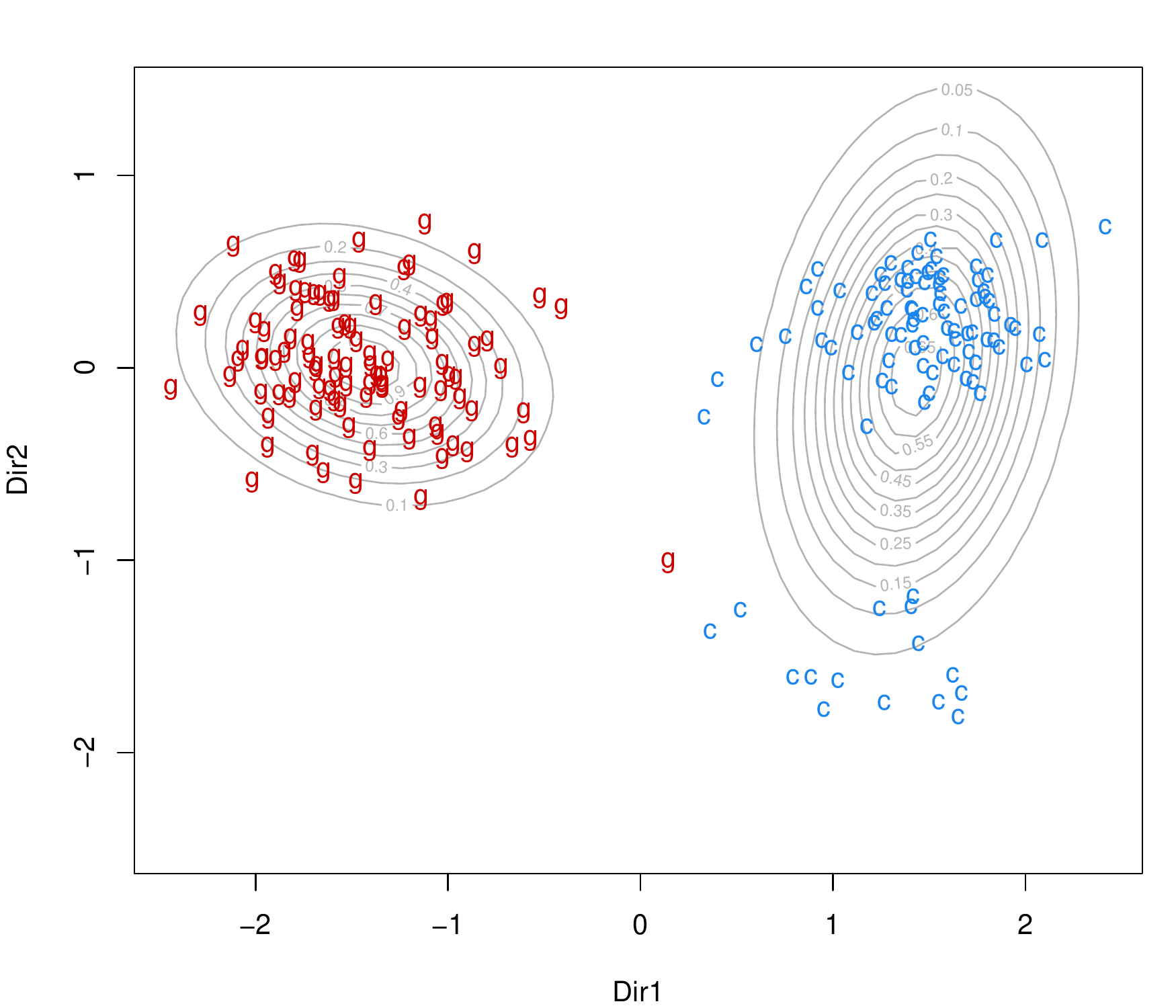}}
\subfloat[]
  {\includegraphics[width = 0.5\linewidth]{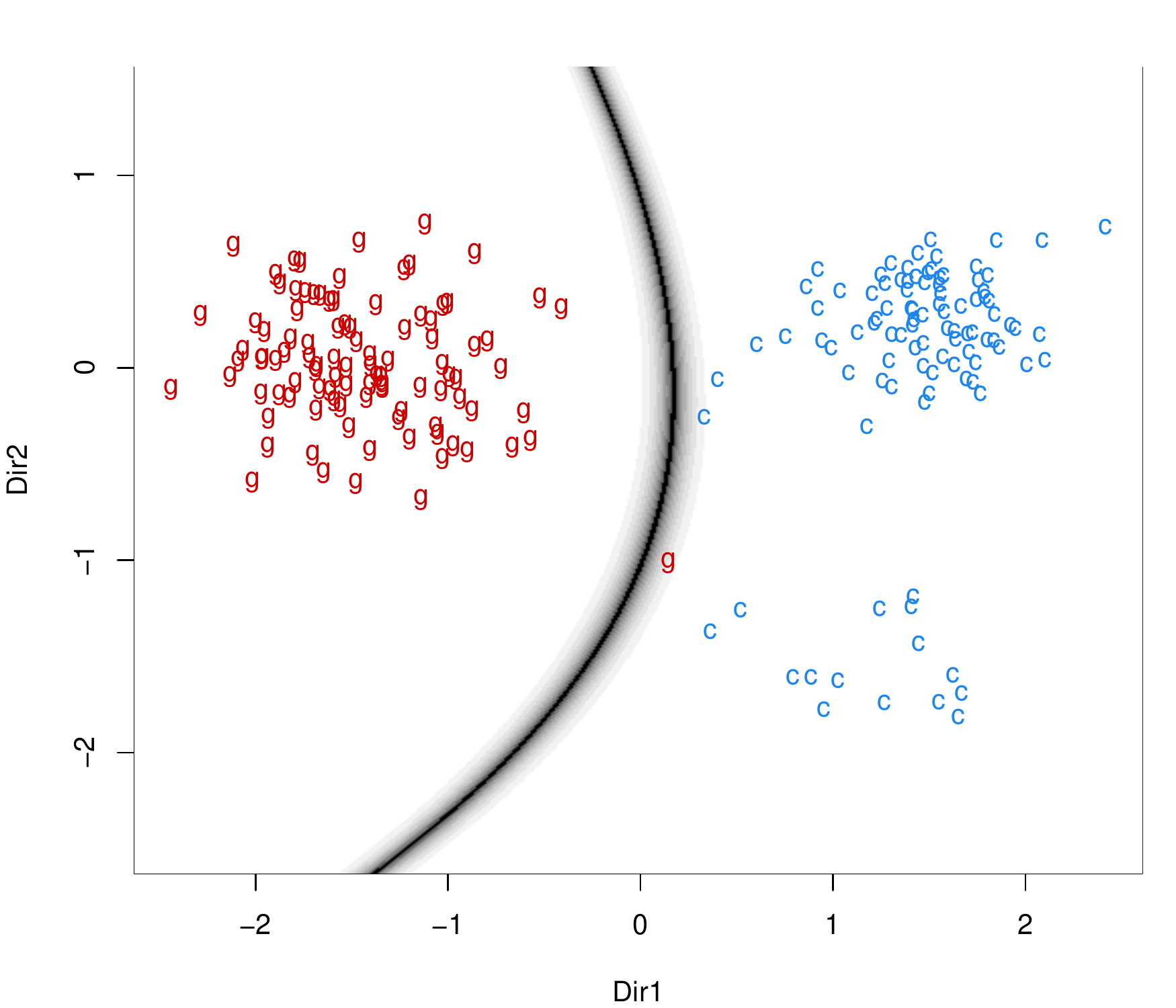}} 
\\[-2ex]
\subfloat[]
  {\includegraphics[width = 0.5\linewidth]{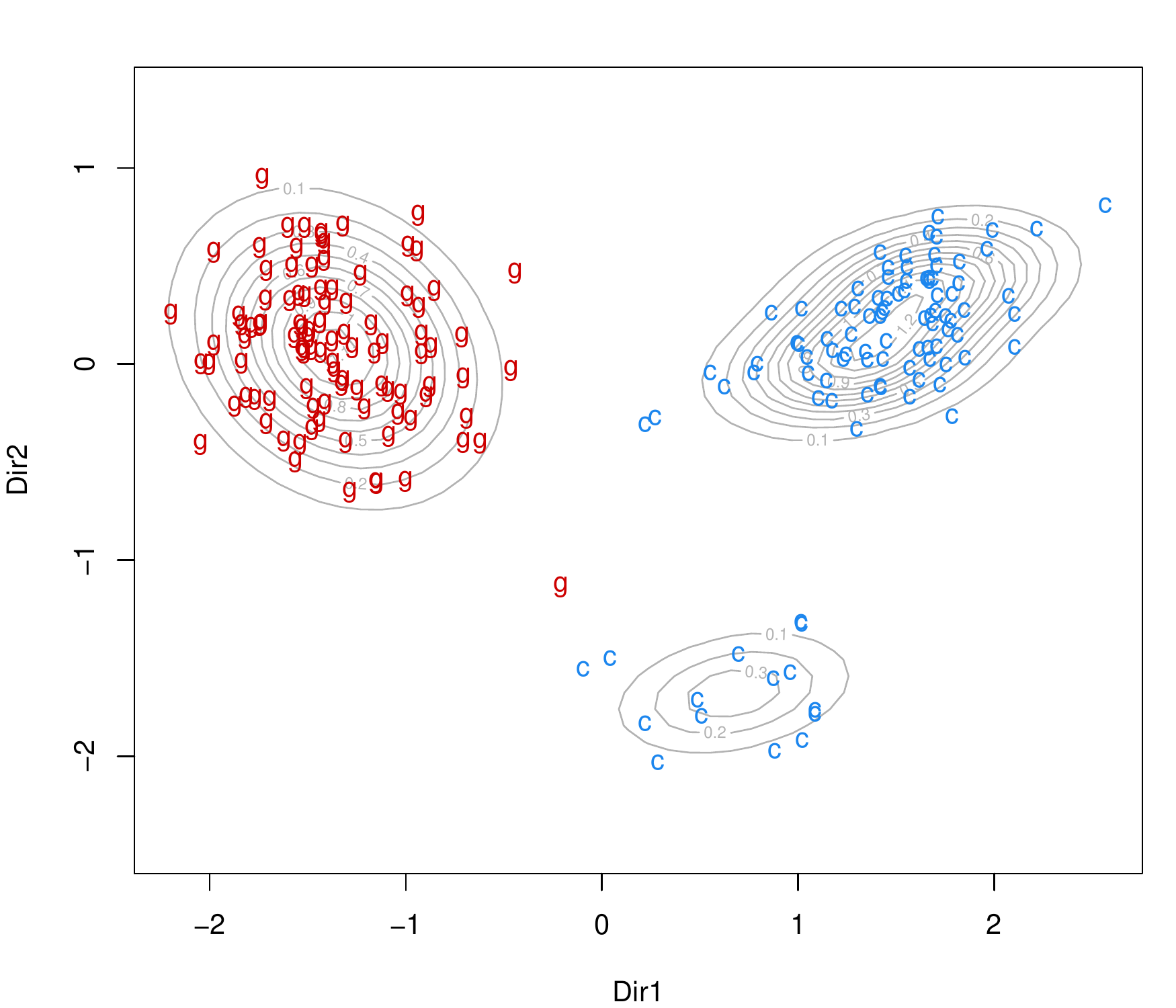}}
\subfloat[]
  {\includegraphics[width = 0.5\linewidth]{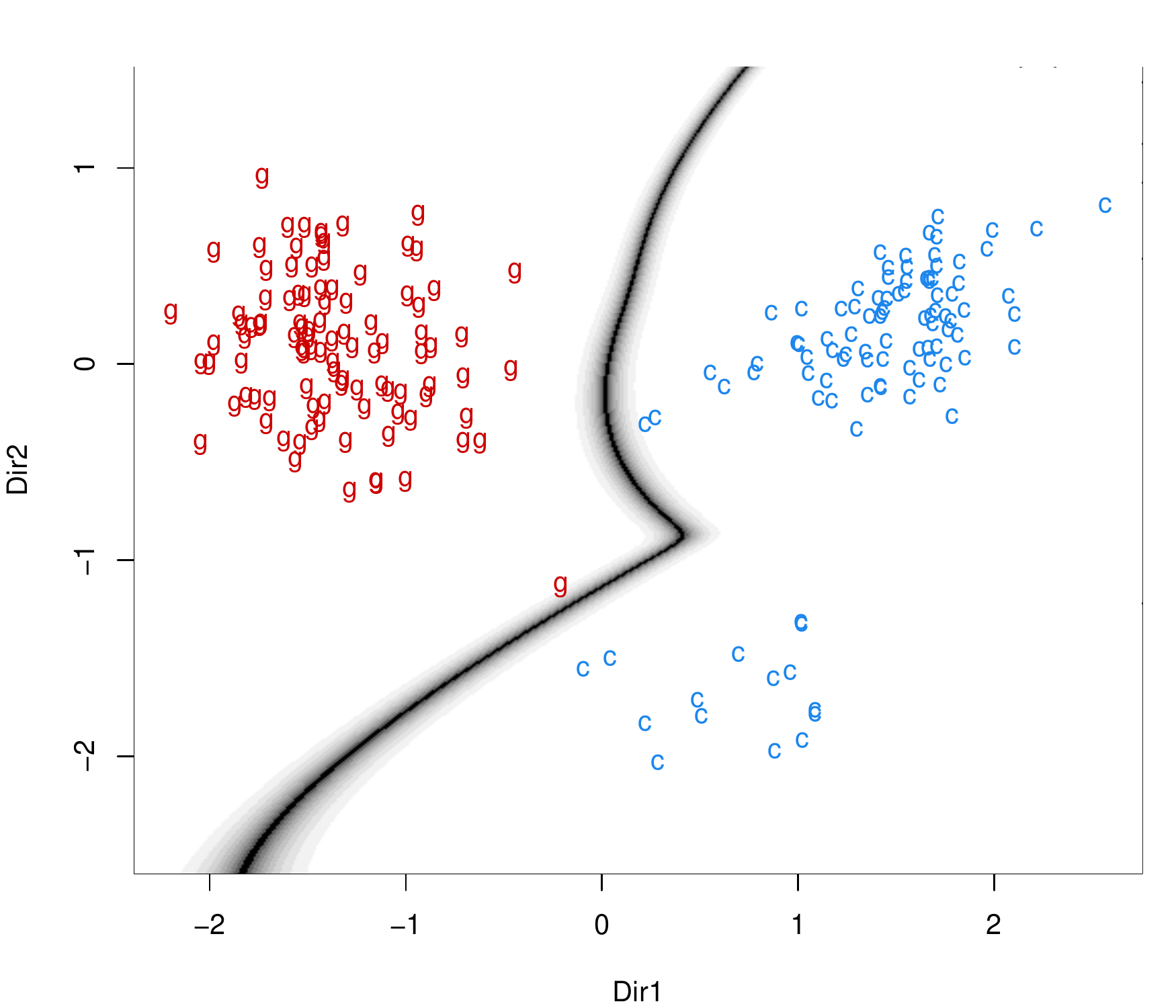}}
\caption{Plot of genuine (\textcolor{red}{g}) and counterfeit (\textcolor{blue}{c}) Swiss banknotes projected onto the first two directions estimated for the ``best'' EDDA mixture model (top panels) and the ``best'' MclustDA mixture model (bottom panels). 
Panels (a) and (c) show the class density contours, while panels (b) and (d) plot the decision boundaries with corresponding uncertainty.}
\label{fig1:banknote}
\end{figure}

The EDDA mixture model selected by BIC is a EEV model, which assume class covariance structures with different orientations but same volume and shape (see Table~\ref{tab:mclust}). Figure~\ref{fig1:banknote}a shows the data projected onto the first two GMMDRC directions with the corresponding density contours; there appears a clear separation between classes with a larger variability for the group of counterfeit banknotes. The corresponding classification boundary is quadratic with an outlying genuine note classified as counterfeit (see panel (b) of Figure~\ref{fig1:banknote}). The estimated subspace is quite similar to that obtained by SAVE, which we recall is equivalent to the one we would have obtained by fitting an EDDA mixture model with VVV covariance structure.

Fitting a MclustDA mixture model we obtain the graphs in panels (c) and (d) of Figure~\eqref{fig1:banknote}. 
The model selected by BIC uses a three components mixture with common covariance structure for the group of counterfeits, and a single component mixture model for the group of genuine notes. The latter appears as an homogeneous group, whereas the counterfeits are more heterogeneous with a clear separated cluster of observations (see panel (c)). Finally, panel (d) of Figure~\eqref{fig1:banknote} shows the classification boundaries which are clearly nonlinear in this case and classify correctly all the observed banknotes.

\subsection{Simulated data with irrelevant and redundant features}

GMMDRC directions are able to identify those variables which contain information about the classification structure. 
The estimated basis of the subspace is thus formed by linear combinations of the original features. However, when irrelevant and/or redundant correlated variables are present, the corresponding estimated coefficients have negligible values.

Consider the synthetic data example described in \citet[][Sec. 6, scenario 5]{Maugis:etal:2009a}. A sample of size $n=200$ is generated for a 10-dimensional ifeature vector. The first two variables are drawn from a mixture of four Gaussian distributions $\x_{[1:2]} \sim N(\mub_k, \I_2)$ with $\mub_1=(-2,-2)$, $\mub_2=(-2,2)$, $\mub_3 = -\mub_2$, $\mub_4 = -\mub_1$, and mixing probabilities $\pi = (0.3,0.2,0.3,0.2)$. The remaining eight variables are simulated according to the model $\x_{[3:10]} = \betab\T\x_{[1:2]} + \epsilon$, where $\betab = \left(\begin{matrix} 0.5 & 0 & 2 & 0\\ 0 & 1 & 0 & 3\end{matrix}\;\0_4\right)$, $\epsilon \sim N(\0_{10}, \Omegab)$ with $\Omegab = \diag(\I_2, 0.5\I_2, \I_4)$, and $\0_p$ is the $p \times p$ matrix of zeroes. 
For this scenario only the first two variables contain relevant information for classification; 
the following four variables are correlated with the first two and therefore are redundant for classification purposes, whereas the remaining variables are independent both from the previous variables and from the classification.

\begin{figure}[htb]%
\begin{minipage}[c]{0.6\linewidth}%
  \includegraphics[width = \linewidth]{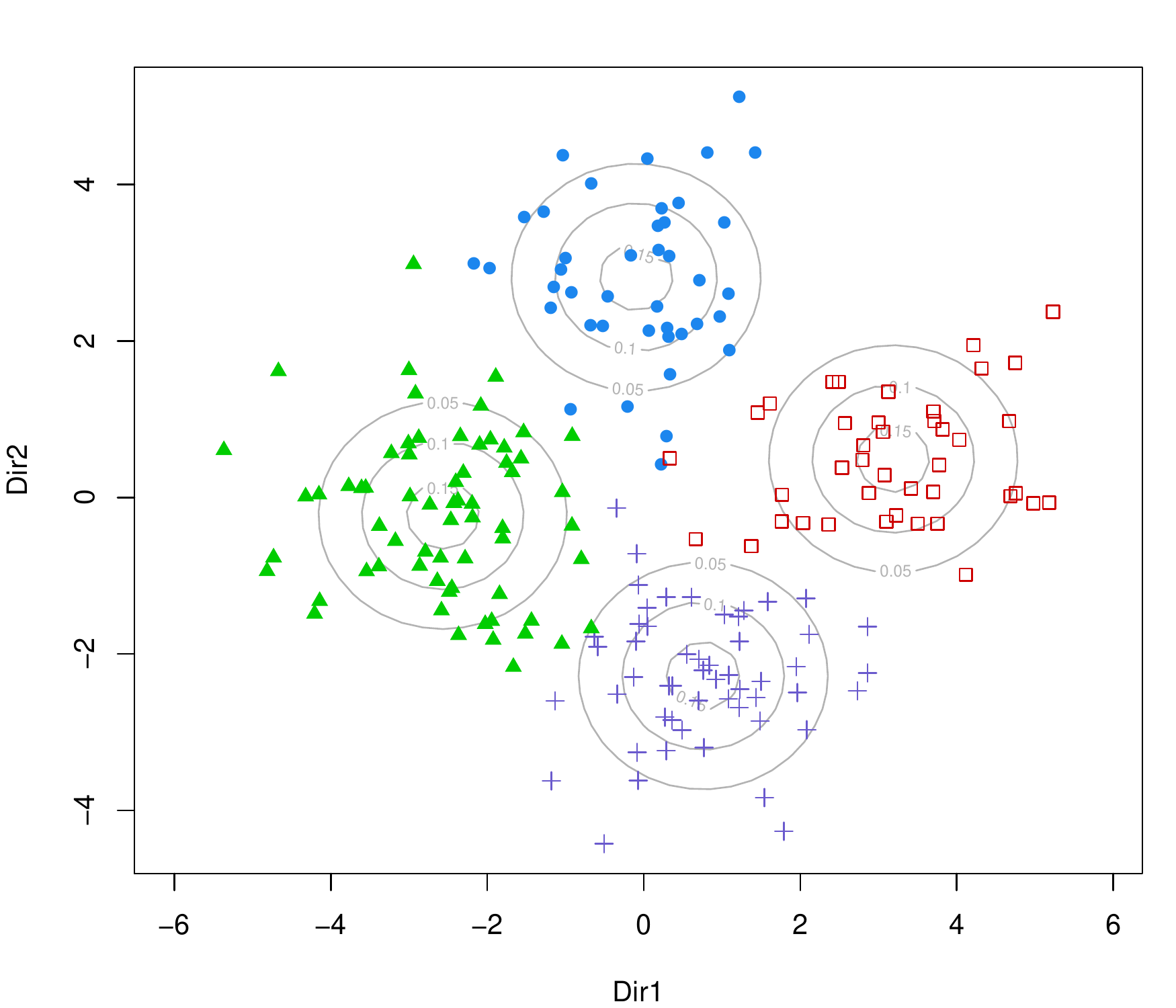}
\end{minipage}%
\quad
\begin{minipage}[c]{0.4\linewidth}%
  \setlength{\tabcolsep}{0.5ex}
  \begin{tabular}{lrrr}
  \multicolumn{4}{c}{GMMDRC basis}\\
  \hline
   & Dir$_1$ & Dir$_2$ & Dir$_3$ \\ 
  \hline
  $x_1$    & $-0.729$ & $-0.481$ & $-0.368$ \\ 
  $x_2$    & $0.672$ & $-0.843$ & $0.701$ \\ 
  $x_3$    & $-0.008$ & $0.012$ & $-0.169$ \\ 
  $x_4$    & $0.051$ & $0.055$ & $0.123$ \\ 
  $x_5$    & $0.052$ & $-0.138$ & $0.212$ \\ 
  $x_6$    & $0.046$ & $0.083$ & $-0.270$ \\ 
  $x_7$    & $-0.023$ & $-0.025$ & $-0.183$ \\ 
  $x_8$    & $0.033$ & $-0.087$ & $-0.363$ \\ 
  $x_9$    & $0.011$ & $-0.035$ & $0.033$ \\ 
  $x_{10}$ & $0.083$ & $0.143$ & $0.214$ \\ 
  \hline
  Eigenvalues & $0.6527$ & $0.6069$ & $0.0005$ \\ 
  Cum. \% & $51.79$ & $99.96$ & $100.00$ \\ 
  \hline
  \end{tabular}
\end{minipage}%
\caption{Plot of simulated data, generated with irrelevant and redundant variables, projected onto the subspace spanned by the first two GMMDRC directions. The table at right shows the coefficients of the linear combinations that define the estimated directions with the corresponding eigenvalues.}%
\label{fig1:ircv}%
\end{figure}

The plot of the sample data projected onto the subspace spanned by the first two GMMDRC directions is presented in Figure~\ref{fig1:ircv}, which also contains the table of coefficients defining the basis of the estimated subspace. 
The first two GMMDRC directions contain all the information pertaining to the partition of the classes, with the last direction clearly negligible based on the value of the corresponding eigenvalue.
Furthermore, the coefficients defining the first two directions are very close to zero for all the variables except the first two, those which are really needed for classification.

\subsection{High-dimensional data example} 

High-dimensional data represents a very challenging problem for many statistical methods, particularly when the number of available observations is small compared to the number of variables. Finite mixture models may be highly parameterized, thus fitting Gaussian mixtures to high-dimensional data requires some form of dimension reduction and/or some form of regularization. For a recent review see \cite{Bouveyron:Brunet:2012}.

Microarray data are an extreme case of high-dimensional data, for the reason that the number of variables (genes) is usually much larger than the number of observations (samples). For instance, consider the famous gene expression cancer dataset from \cite{Golub:etal:1999}. The data contain information on gene expressions in samples from human acute myeloid (AML) and acute lymphoblastic (ALL) leukemias obtained from high-density Affymetrix oligonucleotide arrays. There are 3571 genes and 38 samples: 27 in class ALL, and 11 in class AML. The samples in class ALL could be further split into B-cell and T-cell types.
A preliminary filtering of genes, based on t-tests with p-values adjusted for multiple comparisons using the \citet{Benjamini:Hochberg:1995} method, selected a subset of 731 genes differentially expressed. 
Then, an MclustDA model was fitted on the selected subset assuming a common spherical covariance matrix (EII) for each component within any class. 
From this model the matrices $\MI$ and $\MII$ can be estimated as discussed in Section~\ref{GMMDRC:estimation}. 
However, to apply the eigen-decomposition \eqref{GMMDRC:optimization} we need a regularized estimate of the marginal covariance matrix. Several approaches could be adopted, but here for simplicity we used $\hat{\Sigmab}_X = \diag[s_{i}^{2}]_{i=1,\ldots,p}$, where $s_{i}^{2}$ is the sample variance of the $i$-th gene. Such estimate ignores correlations between genes, which is not biologically realistic, but it has been shown to have no effect on classification accuracy \citep{Dudoit:Fridllyand:Speed:2002}.

\begin{figure}[htbp]
\centering
\subfloat[]
  {\includegraphics[width = 0.5\linewidth]{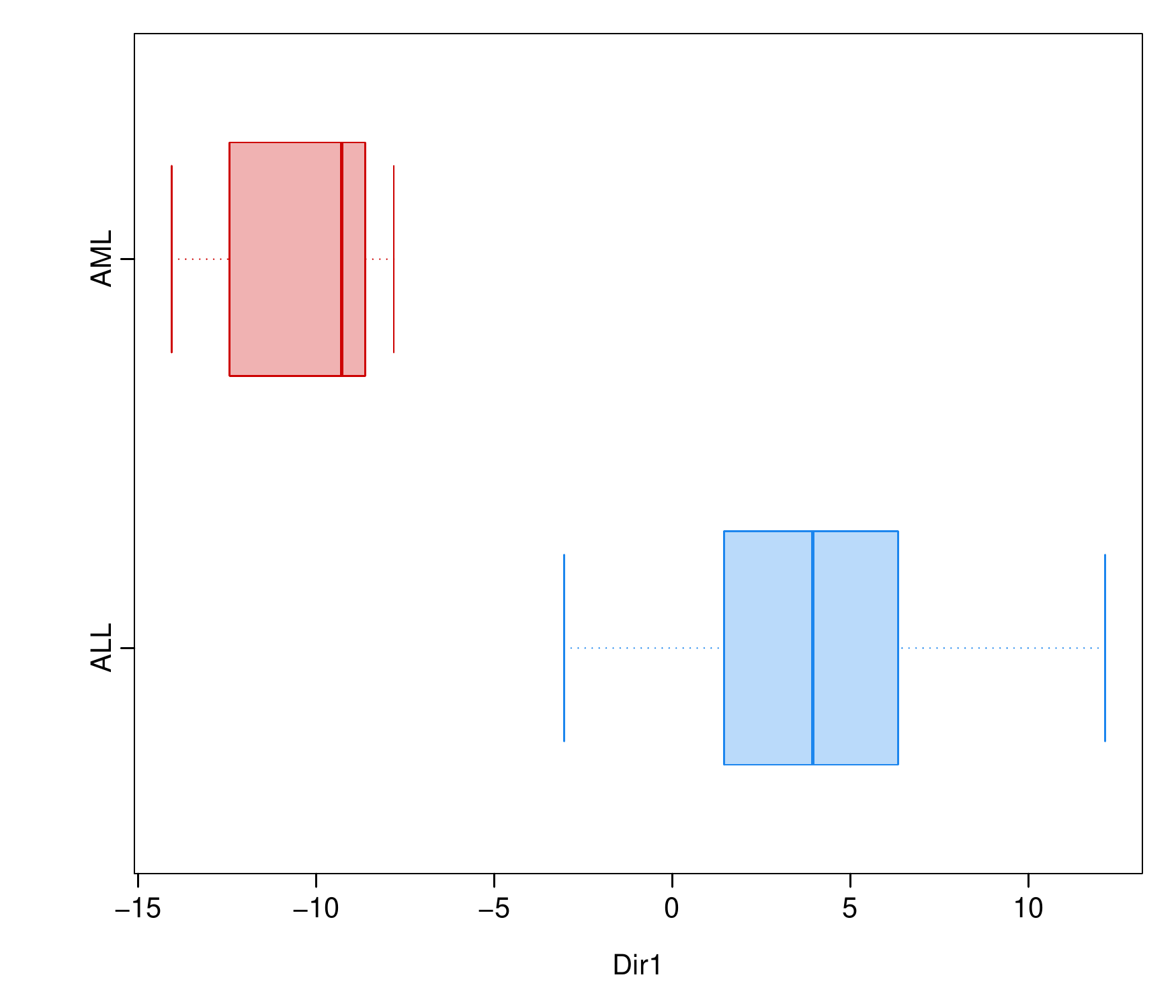}}
\subfloat[]
  {\includegraphics[width = 0.5\linewidth]{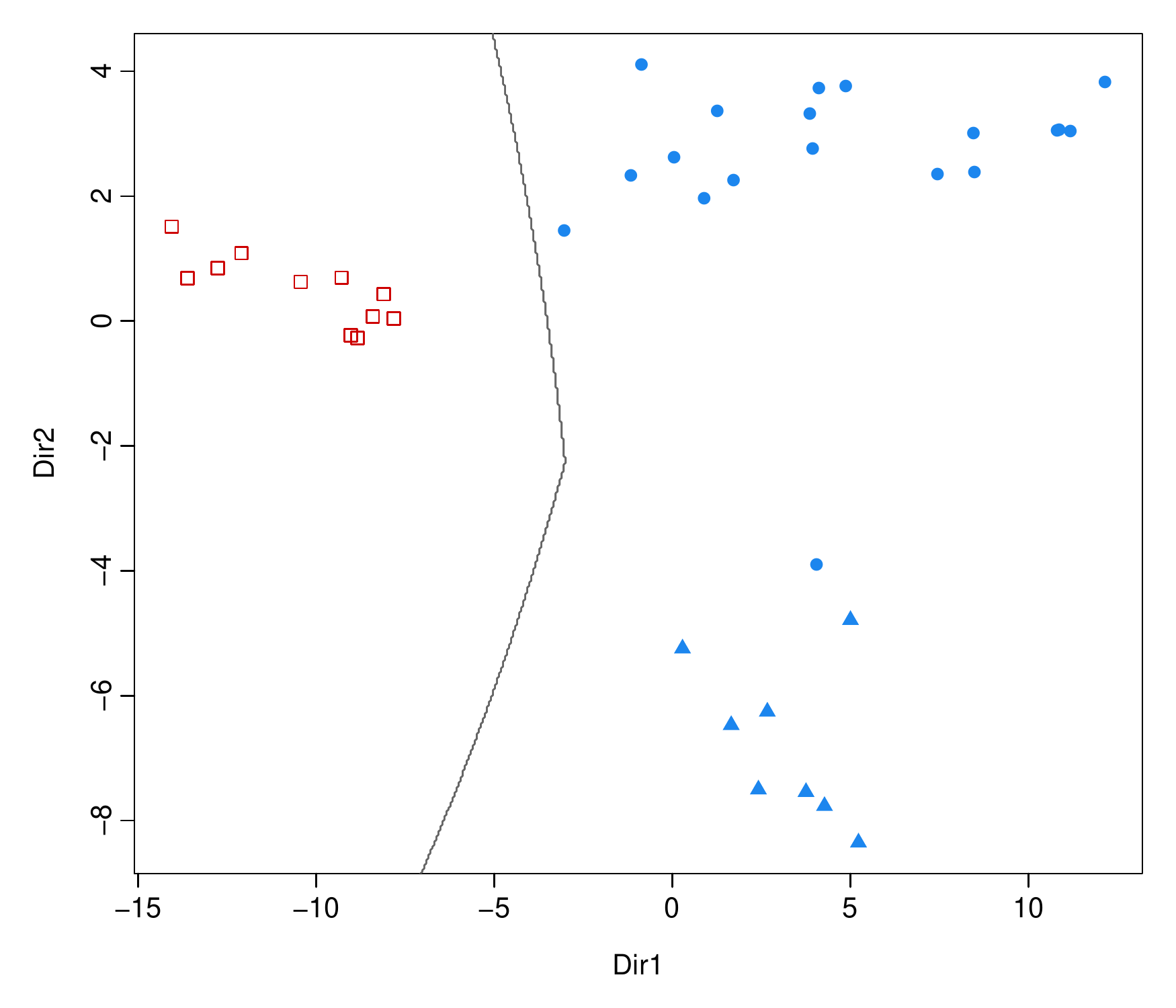}} 
\\
\caption{Plots of Golub data projected along the first GMMDRC direction (a) and the first two GMMDRC directions (b). Points are marked as \textcolor{red}{$\square$} for AML samples, as \textcolor{blue}{\large$\bullet$} for ALL B-cell and \textcolor{blue}{\large$\blacktriangle$} for ALL T-cell samples.}
\label{fig1:golub}
\end{figure}

Figure~\ref{fig1:golub}a shows a boxplot of AML and ALL samples projected along the first GMMDRC direction, which accounts for about 94\% of total variation. A single direction is clearly able to separate the two types of cancer. 
However, the inclusion of the second direction, which accounts for another 4\%, allows us to highlight an interesting feature  previously not evident. Looking at Figure~\ref{fig1:golub}b we see that the group of ALL samples can be further divided into B-cell and T-cell tumour types along the second GMMDRC direction, except for one unusual B-cell which is very close to the group of T-cell samples.

\section{Finding the most discriminant directions}

The methodology introduced in Section~\ref{sec:drmbda} allows to visualize on a reduced subspace the underlying characteristics of class densities.
However, if groups differ not only on location but also on dispersion, this second type of information may be dominant, and the classes would not appear clearly separated along the main directions. 

If class separation is the goal, an appropriate modification of the kernel matrix \eqref{GMMDRC:kernel} should be adopted, for instance, using the following convex linear combination 
\begin{equation}
\M = \lambda\MI \Sigmab^{-1} \MI + (1-\lambda)\MII,
\label{GMMDRC:kernel:lambda}
\end{equation}
where $0 \le \lambda \le 1$ is a tuning parameter. By choosing a large $\lambda$ the estimated directions will focus more on  differences on location. For $\lambda = 0.5$ we give equal weight to the two types of information, while for $\lambda = 1$ differences in class covariances are completely ignored.
More generally, we could decide to optimize a measure of class separation, or minimize the uncertainty in classification.

\begin{figure}[htb]
\centering
\subfloat[]
  {\includegraphics[width = 0.5\linewidth]{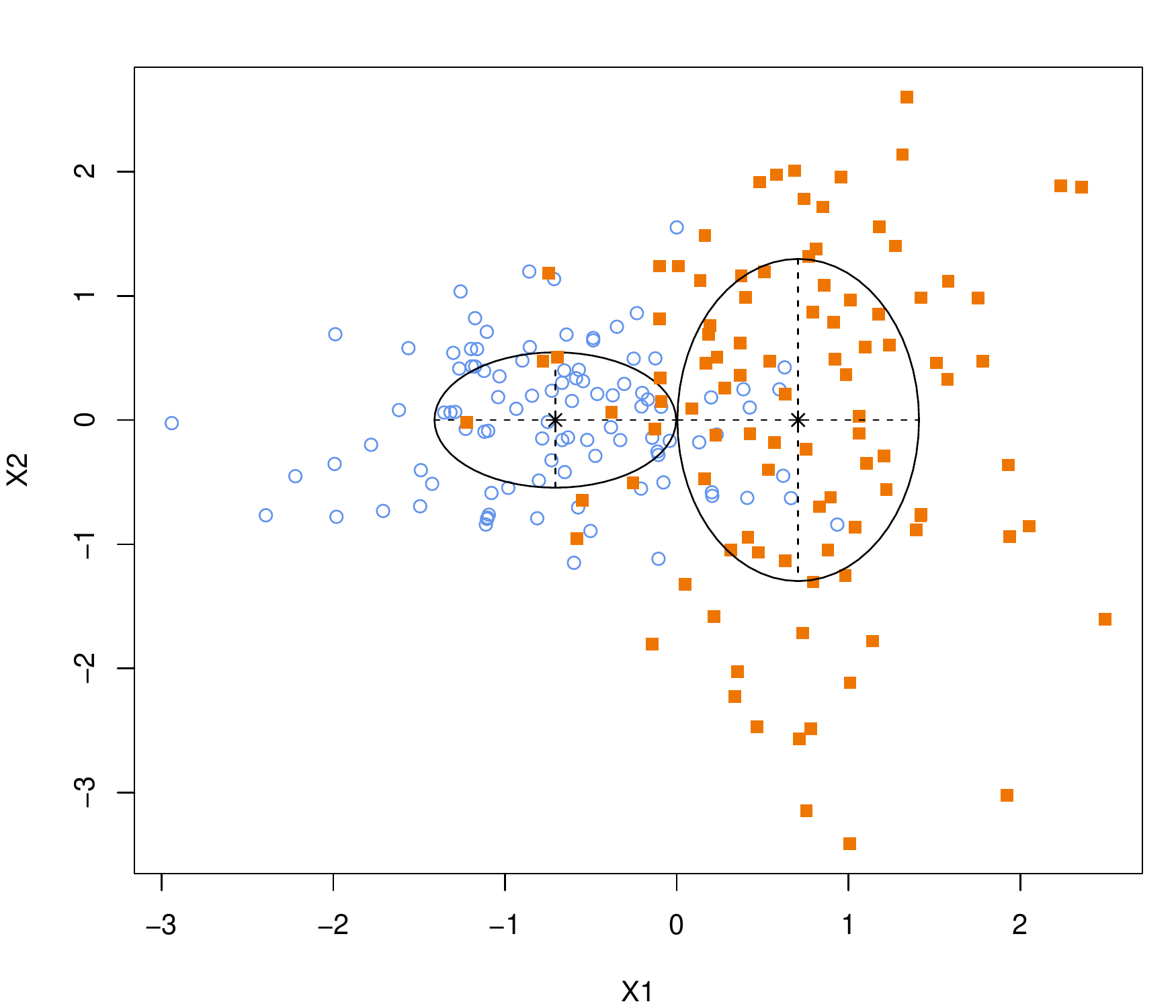}}
\subfloat[]
  {\includegraphics[width = 0.5\linewidth]{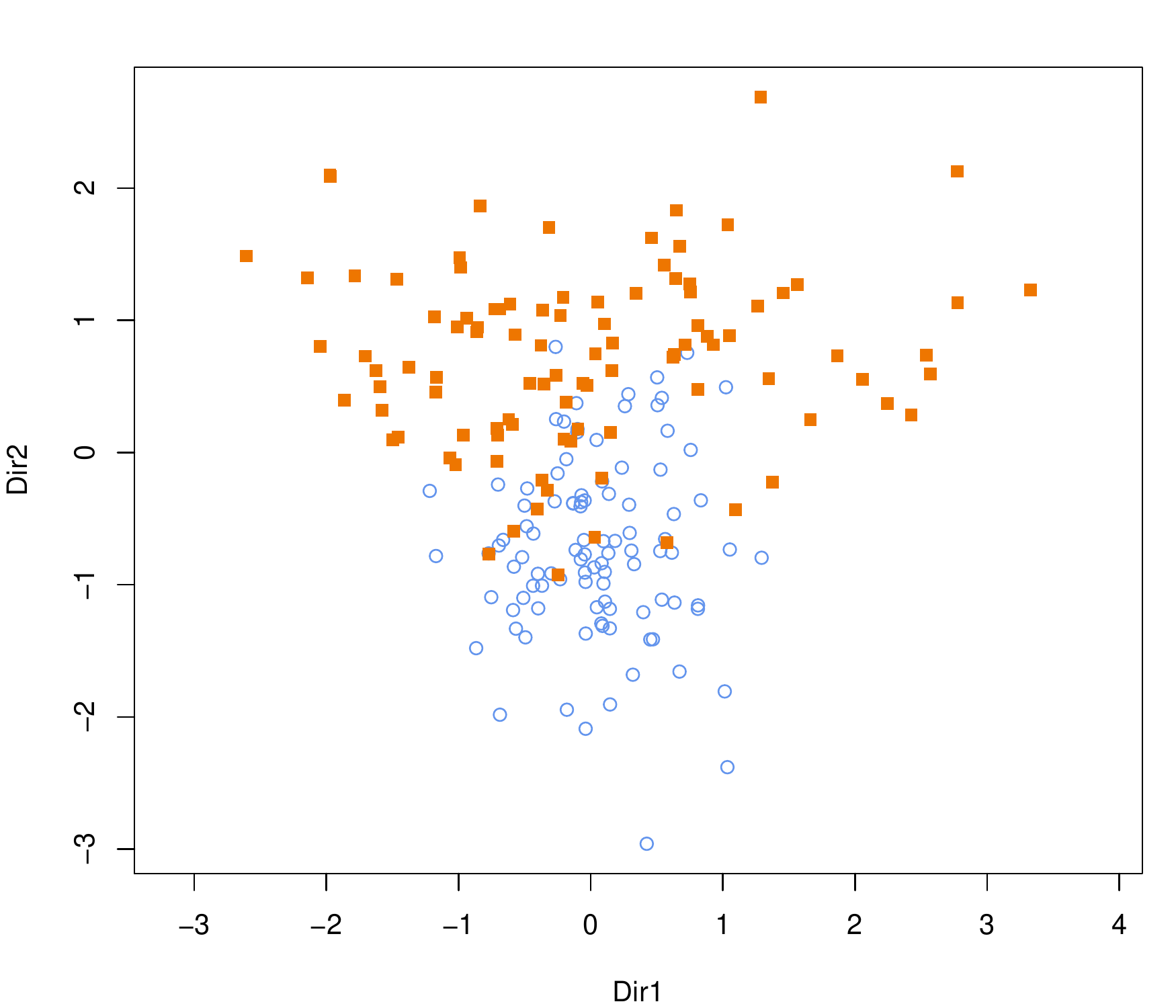}} \\
\subfloat[]
  {\includegraphics[width = 0.5\linewidth]{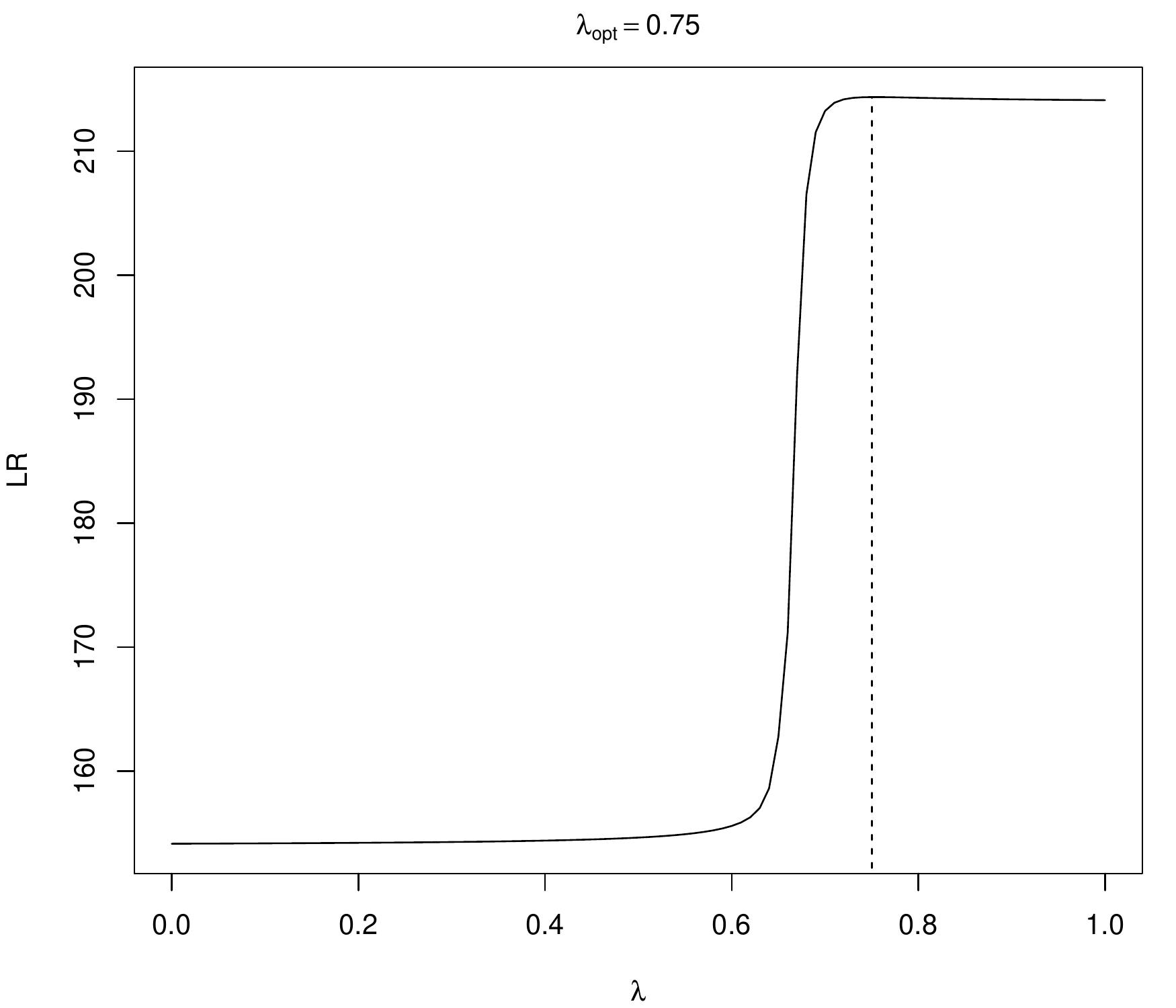}}
\subfloat[]
  {\includegraphics[width = 0.5\linewidth]{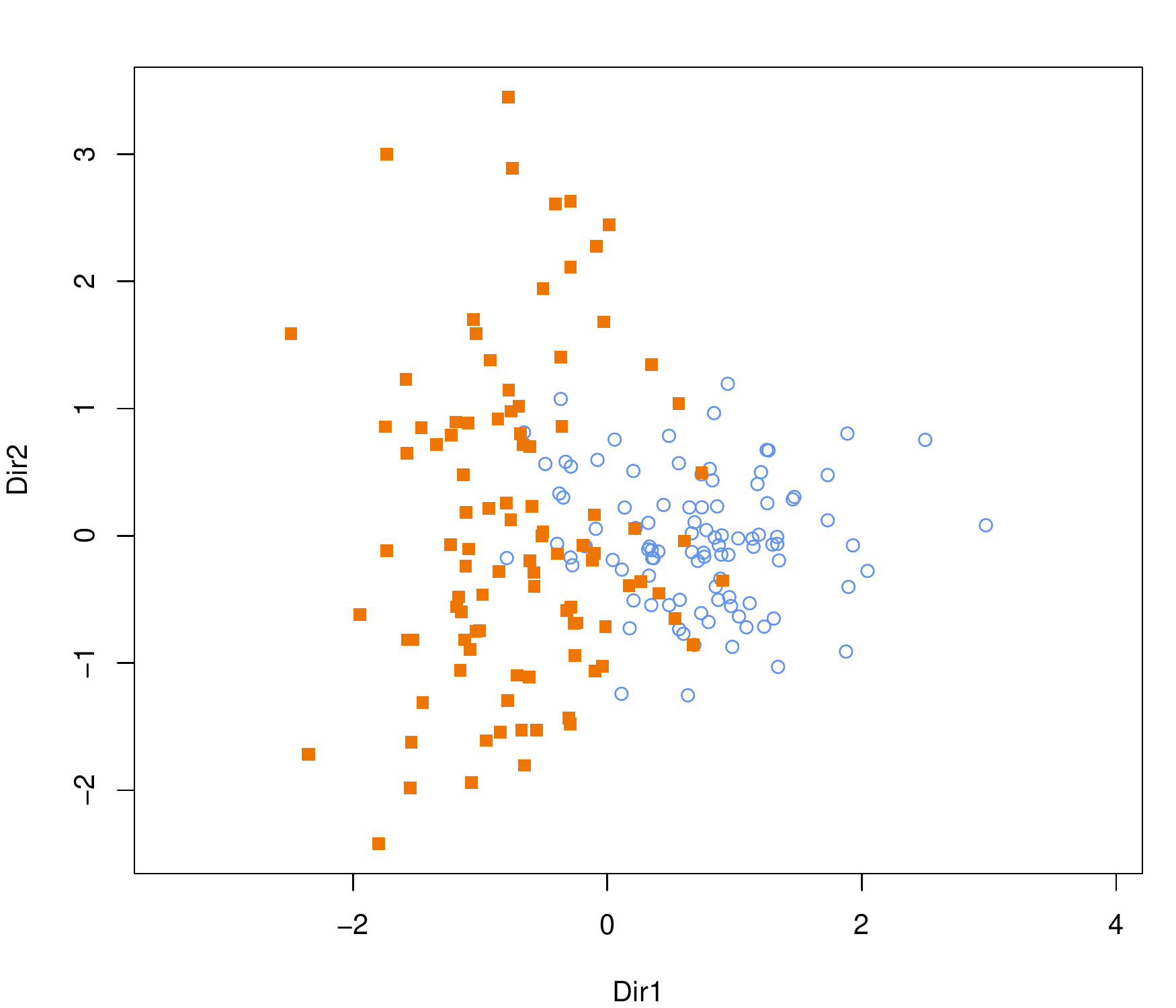}}
\caption{Sample data from \cite{Zhu:Hastie:2003} simulation scheme.  Panel (a) contains the scatterplot of the two classification variables. Panel (b) and (d) show the data projected along the first two estimated GMMDRC directions with, respectively, $\lambda=0.5$ (default) and $\lambda=0.75$. The latter value has been selected on the basis of the LR criterion, whose trace is shown in panel (c).}
\label{fig1:zhu_hastie_ex}
\end{figure}

Recently, \cite{Zhu:Hastie:2003} proposed a generalized log-likelihood-ratio (LR) statistic criterion to find the relevant directions for classification. They compare their proposal with SAVE on a simple bi-dimensional dataset with two groups. Figure~\ref{fig1:zhu_hastie_ex}a shows a data sample generated from this setting (for details see the mentioned paper). \cite{Zhu:Hastie:2003} argue that the relevant direction for discriminating the two groups corresponds to the first variable $X_1$, where there are differences in means. This turns out to be the direction selected by the LR criterion they proposed. On the contrary, the first direction selected by SAVE corresponds to $X_2$, a direction which contains only differences in variances, so the two groups do not appear well separated. 

We fit a Gaussian mixture model to such dataset after adding eight noise variables generated from independent standard normals. The first two directions appears to be needed with associated eigenvalues $(0.54923, 0.28002)$, which accounts for a total contribution of about 83\%. 
Figure~\ref{fig1:zhu_hastie_ex}b shows the data projected along such directions: the first direction correspond essentially to $X_2$, while the second direction to $X_1$. As for SAVE, the information coming from difference in variances is overwhelming that coming from difference in means, and this is what the plot shows. 
However, if our goal is to look for the most separating directions we can adopt the LR criterion for selecting the value of $\lambda$ in equation \eqref{GMMDRC:kernel:lambda}. Figure~\ref{fig1:zhu_hastie_ex}c shows the trace of LR over a regular grid for $\lambda$: the optimal value is obtained for $\lambda = 0.75$ (or greater), which yields the projection shown in Figure~\ref{fig1:zhu_hastie_ex}d. Now, the first estimated direction is essentially equivalent to $X_1$, the most discriminating variable, while the second direction is equivalent to $X_2$.

Instead of optimizing the LR criterion as discussed above, we could adopt a different perspective based on dynamic graphics. For example, one could imagine to build a dynamic graph that, using a slider to manipulate the $\lambda$ parameter, allows us to change interactively the data projection. In this way, a user would be able to appreciate the transition between the focus directed to differences in location to differences in the dispersion.
Furthermore, depending on the purpose of the analysis, by tuning the $\lambda$ parameter we could decide to highlight the structure of the classes and their characteristics, or to favor the separation of classes.


\subsection{Ionosphere data}

The ionosphere data were collected by 16 high-frequency antennas in Goose Bay, Labrador, Canada, and contain information about radar signals returned from the ionosphere. ``Good'' samples are those showing evidence of some type of structure in the ionosphere, while ``bad'' returns are those whose signals pass directly through the ionosphere and show no structure. 
A total of 351 signals were received, 225 were ``good'' returns and the remaining 126 were ``bad'' returns. The signals were processed using a function of 2 attributes for each of 17 pulse numbers that describe the complex electromagnetic signal. There are 34 continuous-valued feature variables, although one is a constant of all zeroes. The dataset is taken from the UCI Machine Learning Repository and it is available in the \textsf{R} package \texttt{mlbench}.

\begin{figure}[htbp]
\centering
\subfloat[]
  {\includegraphics[width = 0.5\linewidth]{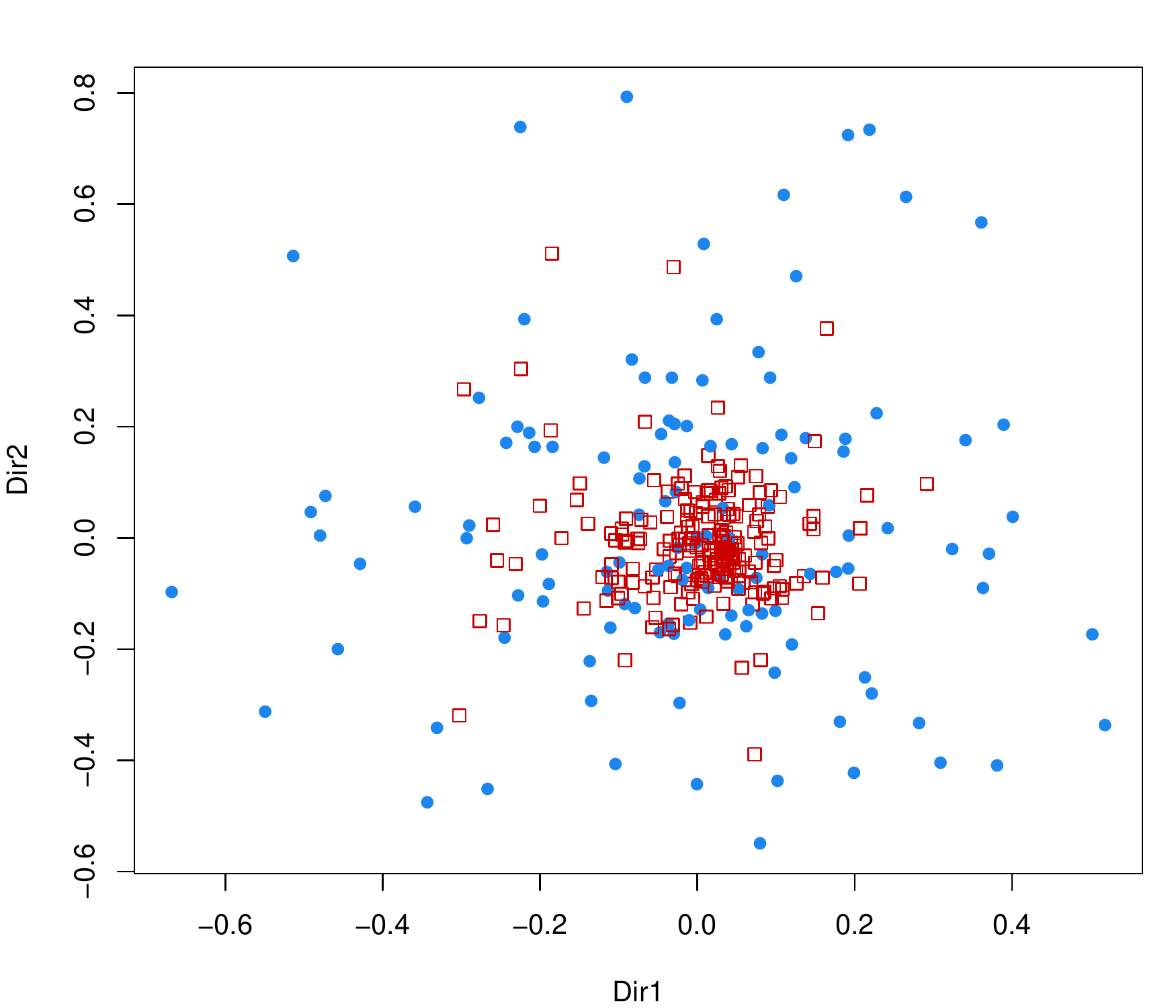}}
\subfloat[]
  {\includegraphics[width = 0.5\linewidth]{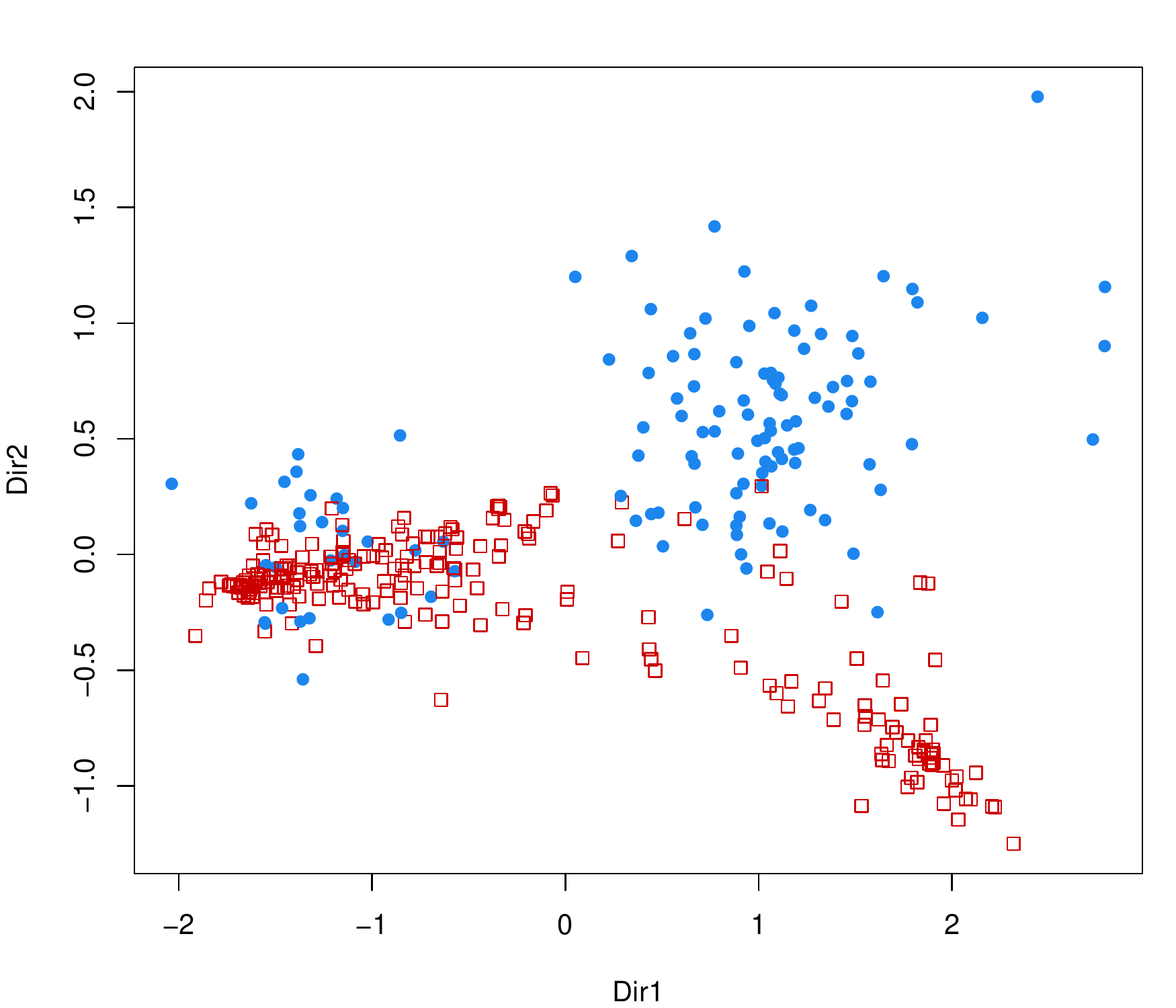}} \\
\caption{Plot of ionosphere data projected onto two different estimated subspaces: (a) using the default $\lambda=0.5$; (b) using the optimal $\lambda = 1$ for groups separation. Points marked as \textcolor{red}{$\square$} refer to ``good'' signals, those marked as \textcolor{blue}{\large$\bullet$} to ``bad'' signals.}
\label{fig1:ionoshere}
\end{figure}

Figure~\ref{fig1:ionoshere}a shows the data projected onto the first two GMMDRC directions using the default $\lambda=0.5$ for a MclustDA mixture model having covariance structure VII with 4 components for the ``bad'' returns, and covariance VVV with 2 components for the ``good'' returns.
On the basis of the graph it can be said that the two groups of signals differ mainly in the dispersion, with the ``bad'' returns showing a larger variance and ``good'' signals which are concentrated around the center of the graph.

These findings are similar to those obtained with SAVE by \cite{Pardoe:Yin:Cook:2007}. To improve groups separation we selected the tuning parameter $\lambda$ using the LR criterion discussed previously. The graph in  Figure~\ref{fig1:ionoshere}b shows the projection onto the first two GMMDRC directions estimated using the optimal value $\lambda=1$. 
Here the separation between the two types of signal is clearly shown along the second direction, while the group of ``good'' signals appears to be composed of two separable sub-groups along the first direction. The latter is an interesting feature not previously recognized.

\section{Final comments}

The paper discussed a dimension reduction method for visualizing the classification structure and the geometric characteristics induced by a Gaussian mixture model. 
The methodology can also be easily adapted in order to recover the directions showing the maximum separation between the classes.

Although in the article we have used two-dimensional projections, the proposed method can, in principle, be easily extended to subspaces of higher dimensions. However, graphical representations on spaces of dimension greater than 3 can be quite difficult. Preliminary results for implementing a guided tour in 2-dimensions seem to be promising.

The methodology and corresponding plots discussed in this paper are available in the \texttt{MclustDR} function of the \textsf{R} package \texttt{mclust} \citep{Fraley:Raftery:Murphy:Scrucca:2012}.




\bibliographystyle{spbasic}
\bibliography{gmmdrc}

\appendix

\subsection*{Proof of proposition~\ref{prop1}}

Assume an EDDA mixture model with common full class covariance matrix. The last condition implies that the matrix $\MII$ in equation \eqref{GMMDRC:kernel} cancels out, so the kernel matrix simplifies to
$$
\M = \MI \Sigmab_X^{-1} \MI,
$$
where $\MI = \sum_{k=1}^K \pi_k (\mub_{k} - \mub)(\mub_{k} - \mub)\T = \Sigmab_B$, the between-class covariance matrix. 
The basis of the subspace $\Space(\betab)$ provided by GMMDRC is obtained as the solution of the following problem
$$
\M \betab_j = l_j \Sigmab_X \betab_j,
$$ 
with $l_1 \ge \ldots \ge l_d$, and $d = \min(p,K-1)$.
Thus, $\betab_j$ is the $j$th eigenvector associated to the $j$th largest eigenvalue $l_j$ ($j=1,\ldots,d$) of the $(p \times p)$ matrix 
\begin{align*}
\Sigmab^{-1/2}_X \M \Sigmab^{-1/2}_X & = 
\Sigmab^{-1/2}_X \MI \Sigmab_X^{-1} \MI \Sigmab^{-1/2}_X \\
& = (\Sigmab^{-1/2}_X \Sigmab_B \Sigmab^{-1/2}_X)\T 
    (\Sigmab^{-1/2}_X \Sigmab_B \Sigmab^{-1/2}_X).
\end{align*}

The subspace estimated by SIR is obtained as the solution of
\begin{equation}
\Sigmab_B \betab_j^\SIR\ = l_j^\SIR \Sigmab_X \betab_j^\SIR
\label{eq:sir:decomp}
\end{equation}
which is given by the eigen-decomposition of $\Sigmab^{-1/2}_X \Sigmab_B\Sigmab^{-1/2}_X$. 
It is easily seen that $\betab_j = \betab_j^\SIR$ and $l_j = (l_j^\SIR)^2$, for $j=1,\ldots,d$.
Thus, the basis of the subspace provided by GMMDRC under model EDDA with full common class covariance matrix is equivalent to the basis estimated by SIR.

We now consider the relation of GMMDRC with LDA canonical variates.
From \eqref{eq:sir:decomp}, we may subtract \;$l^*_j \Sigmab_B \betab_j^\SIR$\; from both side and, recalling the decomposition of the total variance, $\Sigmab_X = \Sigmab_B + \Sigmab_W$, we may write
\begin{align*}
\Sigmab_B \betab_j^\SIR - l_j^\SIR \Sigmab_B \betab_j^\SIR & =  
l_j^\SIR \Sigmab_X \betab_j^\SIR - l_j^\SIR \Sigmab_B \betab_j^\SIR \\
(1 - l_j^\SIR) \Sigmab_B \betab_j^\SIR & = 
l_j^\SIR (\Sigmab_X - \Sigmab_B) \betab_j^\SIR \\
\Sigmab_B \betab_j^\SIR & = 
l_j^\SIR/(1 - l_j^\SIR) \Sigmab_W \betab_j^\SIR.
\end{align*}
It is clear that \;$l_j^\SIR/(1-l_j^\SIR)$\; and \;$\betab_j^\SIR$\; are, respectively, the $j$th eigenvalue and the associated eigenvector of $\Sigmab_W^{-1/2} \Sigmab_B \Sigmab_W^{-1/2}$, the decomposition solving the Rayleigh quotient used to derive canonical variates in LDA. Thus, the basis of the subspace $\Space(\betab^\LDA)$ is equivalent to $\Space(\betab^\SIR)$, which in turn is equivalent to that provided by GMMDRC under the specific model assumption.

\subsection*{Proof of proposition~\ref{prop2}}

The kernel matrix of SAVE can be written in the original scale of the variables as
$$
\M_{\textsf{SAVE}} = \sum_{k=1}^K \w_k
\left( \I_p - \Sigmab_X^{-1/2} \Sigmab_k \Sigmab_X^{-1/2} \right)^2.
$$
Recalling that $\Sigmab_X = \Sigmab_B + \Sigmab_W$, we may write the expression within parenthesis as follows:
\begin{align*}
\Sigmab_X^{-1/2} (\Sigmab_X - \Sigmab_k) \Sigmab_X^{-1/2} & =
\Sigmab_X^{-1/2} (\Sigmab_B + \Sigmab_W - \Sigmab_k) \Sigmab_X^{-1/2} \\
& =
\Sigmab_X^{-1/2} \Sigmab_B \Sigmab_X^{-1/2} + 
\Sigmab_X^{-1/2} (\Sigmab_W - \Sigmab_k) \Sigmab_X^{-1/2}.
\end{align*}
Then,
\begin{align*}
\M_{\textsf{SAVE}} = & \;
\sum_{k=1}^K \w_k
\left( \Sigmab_X^{-1/2} \Sigmab_B \Sigmab_X^{-1/2} + 
       \Sigmab_X^{-1/2}(\Sigmab_W - \Sigmab_k) \Sigmab_X^{-1/2}
\right)^2 \\
= & \;
\Sigmab_X^{-1/2} \Sigmab_B \Sigmab_X^{-1} \Sigmab_B \Sigmab_X^{-1/2} + \\
& \; 
\Sigmab_X^{-1/2} \left(
\sum_{k=1}^K w_k (\Sigmab_k - \Sigmab_W) \Sigmab_X^{-1} (\Sigmab_k - \Sigmab_W)\T 
\right) \Sigmab_X^{-1/2}\\
= & \;
\Sigmab_X^{-1/2} \MI \Sigmab_X^{-1} \MI \Sigmab_X^{-1/2} + \Sigmab_X^{-1/2} \MII \Sigmab_X^{-1/2} \\
= & \;
\Sigmab_X^{-1/2} ( \MI \Sigmab_X^{-1} \MI + \MII ) \Sigmab_X^{-1/2},
\end{align*}
where $\MI$ and $\MII$ are those obtained from an EDDA Gaussian mixture model with a single component for each class and different class covariance matrices (VVV).

\subsection*{Proof of proposition~\ref{prop3}}

The proof is analogous to that provided for Prop. 2 in \cite{Scrucca:2010} and it is not replicated here.

\end{document}